\documentclass{12}

\copyrightauthor{}

\title{The influence of initial solutions to exact solutions of the (2+1)-dimensional generalized Nizhnik-Novikov-Veselov equations}

\author{\footnotesize  Xiao-Feng Yang$^1$$^*$, Zi-Chen Deng$^2$, Qing-Jun Li$^2$ and Yi Wei$^1$\\
$^*$Correspondence:yangxiaofeng@nwsuaf.edu.cn}

\address{1.Department of Applied Mathematics,
Northwestern Polytechnical University, Xi'an 710072, P. R.
China;\\
2.School of Mechanics, Civil Engineering and Architecture,\\
Northwestern Polytechnical University, Xi'an 710072, P. R.
China\\
}

\begin{document}

\maketitle

\begin{abstract}
{\bf Abstract:}\ The (2+1)-dimensional generalized Nizhnik-Novikov-Veselov equations (GNNVEs) are investigated in order to search the influence of initial solution to exact solutions. The GNNVEs are converted into the combined equations of differently two bilinear forms by means of the homogeneous balance of undetermined coefficients method. Accordingly, the two class of exact $N$-soliton solutions and three wave solutions are obtained respectively by using the Hirota's direct method combined with the simplified version of Hereman and the three wave method. The proposed method is also a standard and computable method, which can be generalized to deal with some nonlinear partial differential equations (NLPDEs).\\
{\bf Keywords:}\ generalized Nizhnik-Novikov-Veselov equations; homogeneous balance of undetermined coefficients method; $N$-soliton solution; three wave method\\
{\bf MSC(2010)}: 35A22; 35Q53; 35Q51
\end{abstract}
\pagestyle{myheadings}
 \vskip 0.4 true cm
\section{Introduction}
In recent years, the studies of the exact solutions for NLPDEs have attracted much attention to many mathematicians and physicists [1]. Seeking the exact solutions of NLPDEs has long been an interesting topic in the nonlinear mathematical physics because these exact solutions can describe not only many important phenomena in physics and other fields, but also they can help physicists to understand the mechanisms of the complicated physical phenomena [2-5]. With the development of soliton theory, various methods for obtaining the exact solutions of NLPDEs have been presented, such as the inverse scattering method [6], the homotopy perturbation method [7], the first integral method [8], the variational iteration method [9], the Riccati-Bernoulli sub-ODE method [10], the Jacobi elliptic function method [11], the tanh-sech method [12], the $\left( {G}'/G \right)$-expansion method [13], the Hirota's method [14], the homogeneous balance method [15], the differential transform method [16,17], the Adomian's decomposition method [18-20] and so on.

As is well known, the high dimensional NLPDEs have richer behavior than one-dimensional. It was shown that the different structure of solitons peculiar to the higher dimensionality may contribute to the variety of dynamics of nonlinear waves [6,21,22]. Therefore, the investigation of various structures of exact solutions is imperative for physics and mathematics to the complexity and variety of dynamics determined by higher-dimensional NLPDEs [23,24].

In this paper, we consider the GNNVEs:
$${{u}_{t}}+a{{u}_{xxx}}+b{{u}_{yyy}}+c{{u}_{x}}+d{{u}_{y}}-3a{{\left( uv \right)}_{x}}-3b{{\left( u\omega  \right)}_{y}}=0,\eqno            (1a)$$
$${{u}_{x}}={{v}_{y}},\eqno                            (1b)$$
$${{u}_{y}}={{\omega }_{x}},\eqno                            (1c)$$
which is a known isotropic extension of the well-known (1+1)-dimensional KdV equation, and $a,b,c$ and $d$ are constants. When $c=d=0$, the GNNVEs are reduced to the (2+1)-dimensional Nizhnik-Novikov-Veselov equations (NNVEs) which have been studied by many authors [25-27]. But for the GNNVEs, Radha and Lakshmanan constructed only its dromion solutions from its bilinear form after analyzing its integrability aspects [28], and Zhang et al. obtained its the single solitary wave solutions, multi-soliton solutions and dromion solutions [29]. In this paper, in order to search the influence of initial solution (Obviously, $u={{a}_{000}},v={{b}_{000}},\omega ={{c}_{000}}$ are initial solution of the GNNVEs, where ${{a}_{000}},\ {{b}_{000}}$ and ${{c}_{000}}$ are arbitrary constants.) to the exact solutions, the GNNVEs are converted into the combined equations of differently two bilinear forms by using the homogeneous balance of undetermined coefficients method. Accordingly, the exact $N$-soliton solutions and three wave solutions are easily obtained respectively.

The remainder of this paper is organized as follows: the GNNVEs are converted into the combined equations of differently two bilinear forms in section 2. In sections 3-4, by using the structures which are obtained in section 2, the exact $N$-soliton solutions and three wave solutions are constructed respectively. In section 5, some conclusions are given.

\section{Structure of the GNNVEs}
Now, we deduce the structure of the GNNVEs by using the homogeneous balance of undetermined coefficients method.

Suppose that the solutions of Eqs. (1) are of the forms
$$u={{a}_{mns}}{{\left( \ln w \right)}_{m,n,s}}+\sum\limits_{\begin{smallmatrix}
 \quad i,j,k=0 \\
 i+j\ne 0,m+n+s
\end{smallmatrix}}^{i=m,j=n,k=s}{{{a}_{ijk}}{{\left( \ln w \right)}_{i,j,k}}}+{{a}_{000}},\eqno              (2a)$$
$$v={{b}_{pqg}}{{\left( \ln w \right)}_{p,q,g}}+\sum\limits_{\begin{smallmatrix}
 \quad i,j,k=0 \\
 i+j\ne 0,p+q+g
\end{smallmatrix}}^{i=p,j=q,k=g}{{{b}_{ijk}}{{\left( \ln w \right)}_{i,j,k}}}+{{b}_{000}},\eqno              (2b)$$
$$\omega ={{c}_{lrh}}{{\left( \ln w \right)}_{l,r,h}}+\sum\limits_{\begin{smallmatrix}
 \quad i,j,k=0 \\
 i+j\ne 0,l+r+h
\end{smallmatrix}}^{i=l,j=r,k=h}{{{c}_{ijk}}{{\left( \ln w \right)}_{i,j,k}}}+{{c}_{000}},\eqno               (2c)$$
where $u=u\left( x,y,t \right),\ v=v\left( x,y,t \right),\ \omega =\omega \left( x,y,t \right),\ w=w\left( x,y,t \right)$ and ${{\left(\ln w \right)}_{i,j,k}}=\frac{{{\partial }^{i+j+k}}\left( \ln w\left( x,y,t \right) \right)}{\partial {{x}^{i}}\partial {{y}^{j}}\partial {{t}^{k}}}$, and $m,n,s,p,q,g,l,r,h$ (balance numbers), and ${{a}_{ijk}}\left(i=0,1,\cdots,m;\ j=0,1,\cdots,n;\ k=0,1,\cdots,s\right),\ {{b}_{ijk}}$\\  $\left(i=0,1,\cdots,p;\ j=0,1,\cdots,q;\ k=0,\cdots,g\right)$ and ${{c}_{ijk}}\ \left( i=0,1,\cdots,l;j=0,1,\cdots,r;k=0,1,\cdots,h \right)$\\  $\left( {{a}_{mns}}{{b}_{pqg}}{{c}_{lrh}}\ne 0 \right)$ (balance coefficients) are constants to be determined later.\\
Balancing ${{u}_{xxx}}$ and ${{\left(uv \right)}_{x}}$ in Eq. (1a), ${{u}_{x}}$ and ${{v}_{y}}$ in Eq. (1b), and ${{u}_{y}}$ and ${{\omega }_{x}}$ in Eq. (1c), it is required that\\
$m+p+1=m+3,\ n+q=n,\ s+g=s,\ m+1=p,\ n=q+1,\ s=g,\ m=l+1,\ n+1=r,\ s=h.$\\
Solving the above algebraic equations, we get $m=1,\ n=1,\ s=0,\ p=2,\ q=0,\ g=0,\ l=0,\ r=2,\ h=0$. Then Eqs. (2) can be written as
$$u={{a}_{110}}{{\left( \ln w \right)}_{xy}}+{{a}_{100}}{{\left( \ln w \right)}_{x}}+{{a}_{010}}{{\left( \ln w \right)}_{y}}+{{a}_{000}},\eqno           (3a)$$
$$v={{b}_{200}}{{\left( \ln w \right)}_{xx}}+{{b}_{100}}{{\left( \ln w \right)}_{x}}+{{b}_{000}},\eqno                (3b)$$
$$\omega ={{c}_{020}}{{\left( \ln w \right)}_{yy}}+{{c}_{010}}{{\left( \ln w \right)}_{y}}+{{c}_{000}},\eqno                (3c)$$
where ${{a}_{ij0}}\ \left( i,j=0,1 \right)$, ${{b}_{i00}}\ \left( i=0,1,2 \right)$ and ${{c}_{0i0}}\ \left( i=0,1,2 \right)$ are constants to be determined later.\\
Substituting Eqs. (3) into Eq. (1b) and equating the coefficients of $\frac{w_{x}^{2}{{w}_{y}}}{{{w}^{3}}}$ on the left-hand side of Eqs. (1b) to zero yield an algebraic equation for ${{a}_{110}}$ and ${{b}_{200}}$ as follows:
$$2\left( {{a}_{110}}-{{b}_{200}} \right)=0.\eqno                         (4)$$
Substituting Eqs. (3) into Eq. (1c) and equating the coefficients of $\frac{{{w}_{x}}w_{y}^{2}}{{{w}^{3}}}$ on the left-hand side of Eq. (1c) to zero yield an algebraic equation for ${{a}_{110}}$ and ${{c}_{020}}$ as follows:
$$2\left( {{a}_{110}}-{{c}_{020}} \right)=0.\eqno                        (5)$$
Solving Eqs. (4) and (5), we get ${{b}_{200}}={{c}_{020}}={{a}_{110}}$. Substituting ${{b}_{200}}={{c}_{020}}={{a}_{110}}$ back into Eq. (1a) and equating the coefficients of $\frac{{{w}_{x}}{{w}_{y}}\left( aw_{x}^{3}+bw_{y}^{3} \right)}{{{w}^{5}}}$ on the left-hand side of Eq. (1a) to zero yield an algebraic equation for ${{a}_{110}}$ as follows:
$$12{{a}_{110}}\left( {{a}_{110}}+2 \right)=0.\eqno                         (6)$$
Solving the above algebraic equation and noticing ${{a}_{110}}\ne 0$, we get ${{a}_{110}}=-2$. Substituting ${{a}_{110}}={{b}_{200}}={{c}_{020}}=-2$ back into Eq. (1b) and integrating with respect to $x$ once (taking the integration constant as zero), we get
$$\frac{{{a}_{100}}{{w}_{x}}+\left( {{a}_{010}}-{{b}_{100}} \right){{w}_{y}}}{w}=0.\eqno                       (7)$$
Substituting ${{a}_{110}}={{b}_{200}}={{c}_{020}}=-2$ back into Eq. (1c) and integrating with respect to $y$ once (taking the integration constant as zero), we get
$$\frac{{{a}_{010}}{{w}_{y}}+\left( {{a}_{100}}-{{c}_{010}} \right){{w}_{x}}}{w}=0.\eqno                       (8)$$
Obviously, setting ${{a}_{100}}={{a}_{010}}={{b}_{100}}={{c}_{010}}=0$, Eqs. (1b) and (1c) become identities. Accordingly, Eqs. (2) are reduces to
$$u=-2{{\left( \ln w \right)}_{xy}}+{{a}_{000}},\ v=-2{{\left( \ln w \right)}_{xx}}+{{b}_{000}},\ \omega =-2{{\left( \ln w \right)}_{yy}}+{{c}_{000}},\eqno      (9)$$
where $w=w\left( x,y,t \right)$ is a function of $x,y,t$ that will be determined later, and ${{a}_{000}}$, ${{b}_{000}}$ and ${{c}_{000}}$ are constants to be determined later.\\
Substituting Eqs. (9) into Eq. (1a) and simplifying it, we get
\begin{displaymath}
\begin{aligned}
   {{\left( \frac{2a\left( {{K}_{1112}}-3{{a}_{000}}{{K}_{11}}-3{{b}_{000}}{{K}_{12}} \right)+2c{{K}_{12}}+2d{{K}_{22}}+2{{K}_{23}}}{{{w}^{2}}} \right)}_{x}} \\
  -{{\left( \frac{2b\left( 3{{a}_{000}}{{K}_{22}}+3{{c}_{000}}{{K}_{12}}-{{K}_{1222}} \right)}{{{w}^{2}}} \right)}_{y}}=0, \\
\end{aligned} \eqno(10)
\end{displaymath}
where\\
$${{K}_{11}}={{w}_{xx}}w-w_{x}^{2},\ {{K}_{12}}={{w}_{xy}}w-{{w}_{x}}{{w}_{y}},\ {{K}_{22}}={{w}_{yy}}w-w_{y}^{2},\ {{K}_{23}}={{w}_{yt}}w-{{w}_{y}}{{w}_{t}},$$
$${{K}_{1112}}={{w}_{xxxy}}w+3{{w}_{xx}}{{w}_{xy}}-{{w}_{xxx}}{{w}_{y}}-3{{w}_{x}}{{w}_{xxy}},\ {{K}_{1222}}={{w}_{xyyy}}w-3{{w}_{xyy}}{{w}_{y}}-{{w}_{x}}{{w}_{yyy}}+3{{w}_{xy}}{{w}_{yy}}.$$
Eq. (10) can be written concisely in terms of $D$-operator as
\begin{displaymath}
\begin{aligned}
{{\left( \frac{\left( {{D}_{y}}{{D}_{t}}+aD_{x}^{3}{{D}_{y}}-3a{{a}_{000}}D_{x}^{2}+\left( c-3a{{b}_{000}} \right){{D}_{x}}{{D}_{y}}+dD_{y}^{2} \right)w\cdot w}{{{w}^{2}}} \right)}_{x}} \\
  -b{{\left( \frac{\left( 3{{a}_{000}}D_{y}^{2}+3{{c}_{000}}{{D}_{x}}{{D}_{y}}-{{D}_{x}}D_{y}^{3} \right)w\cdot w}{{{w}^{2}}} \right)}_{y}}=0, \\
\end{aligned} \eqno(11)
\end{displaymath}
where
$${{\left. D_{x}^{m}D_{t}^{n}a\cdot b={{\left( {{\partial }_{x}}-{{\partial }_{{{x}'}}} \right)}^{m}}{{\left( {{\partial }_{t}}-{{\partial }_{{{t}'}}} \right)}^{n}}a\left( x,t \right)b\left( {x}',{t}' \right) \right|}_{{x}'=x,{t}'=t}}.$$
From Eq. (11), we suppose
$$\left( {{D}_{y}}{{D}_{t}}+aD_{x}^{3}{{D}_{y}}-3a{{a}_{000}}D_{x}^{2}+\left( c-3a{{b}_{000}} \right){{D}_{x}}{{D}_{y}}+dD_{y}^{2} \right)w\cdot w=0,\eqno       (12)$$
and
$$\left( 3{{a}_{000}}D_{y}^{2}+3{{c}_{000}}{{D}_{x}}{{D}_{y}}-{{D}_{x}}D_{y}^{3} \right)w\cdot w=0.\eqno               (13)$$
Eqs. (12) and (13) are the combined equations of differently two bilinear forms for the GNNVEs.

\section{$N$-soliton solutions of the GNNVEs}
In this section, to obtain $N$-soliton solutions for the GNNVEs, we will apply the Hirota's direct method [14] combined with the simplified version of Hereman et al. [30,31] where it was shown that soliton solutions are just polynomials of exponentials. Moreover, we will show that $N$-soliton solutions for finite $N\ \left( N\ge 1 \right)$ exist [32].

To determine the dispersion relation, we set ${{c}_{000}}=0$ and
$$w=1+{{e}^{Px+\Omega y+Kt}}.\eqno                           (14)$$
Substituting Eq. (14) into Eqs. (12) and (13) yields the dispersion relations
$$\Omega =\frac{3{{a}_{000}}}{P},\ K=\frac{3{{P}^{2}}a{{b}_{000}}-c{{P}^{2}}-3d{{a}_{000}}}{P}.$$
Accordingly, we get the 1-soliton solution, 2-soliton solution, 3-soliton solution as follows:
$${{w}_{1}}=1+{{e}^{{{\eta }_{1}}}},\ {{w}_{2}}=1+{{e}^{{{\eta }_{1}}}}+{{e}^{{{\eta }_{2}}}}+{{a}_{12}}{{e}^{{{\eta }_{1}}+{{\eta }_{2}}}},$$
$${{w}_{3}}=1+{{e}^{{{\eta }_{1}}}}+{{e}^{{{\eta }_{2}}}}+{{e}^{{{\eta }_{3}}}}+{{a}_{12}}{{e}^{{{\eta }_{1}}+{{\eta }_{2}}}}+{{a}_{13}}{{e}^{{{\eta }_{1}}+{{\eta }_{3}}}}+{{a}_{23}}{{e}^{{{\eta }_{2}}+{{\eta }_{3}}}}+{{a}_{123}}{{e}^{{{\eta }_{1}}+{{\eta }_{2}}+{{\eta }_{3}}}},\eqno      (15)$$
$${{u}_{i}}=-2{{\left( \ln {{w}_{i}} \right)}_{xy}}+{{a}_{000}},\ {{v}_{i}}=-2{{\left( \ln {{w}_{i}} \right)}_{xx}}+{{b}_{000}},\ {{\omega }_{i}}=-2{{\left( \ln {{w}_{i}} \right)}_{yy}},$$
where
$${{\eta }_{i}}={{K}_{i}}t+{{\Omega }_{i}}y+{{P}_{i}}x,\ {{\Omega }_{i}}=\frac{3{{a}_{000}}}{{{P}_{i}}},\  {{K}_{i}}=\frac{3P_{i}^{2}a{{b}_{000}}-cP_{i}^{2}-3d{{a}_{000}}}{{{P}_{i}}},$$
$${{a}_{ij}}=\frac{\left( P_{i}^{2}-{{P}_{i}}{{P}_{j}}+P_{j}^{2} \right){{\left( {{P}_{i}}-{{P}_{j}} \right)}^{2}}}{\left( P_{i}^{2}+{{P}_{i}}{{P}_{j}}+P_{j}^{2} \right){{\left( {{P}_{i}}+{{P}_{j}} \right)}^{2}}},\ \left( i,j=1,2,3;i<j \right),\ {{a}_{123}}={{a}_{12}}{{a}_{13}}{{a}_{23}},$$
and ${{P}_{i}}\ \left( {{P}_{i}}\ne 0 \right)\ \left( i=1,2,3 \right),\ {{a}_{000}}$ and ${{b}_{000}}$ are arbitrary constants.\\
Notice that we use Eq. (14) to determine the dispersion relation, ${{w}_{2}}=1+{{e}^{{{\eta }_{1}}}}+{{e}^{{{\eta }_{2}}}}+{{a}_{12}}{{e}^{{{\eta }_{1}}+{{\eta }_{2}}}}$ to determine the factor ${{a}_{12}}$ to generalize the result for the other factors ${{a}_{ij}}\ \left( 1\le i<j\le N \right)$, and finally we use ${{w}_{3}}=1+{{e}^{{{\eta }_{1}}}}+{{e}^{{{\eta }_{2}}}}+{{e}^{{{\eta }_{3}}}}+{{a}_{12}}{{e}^{{{\eta }_{1}}+{{\eta }_{2}}}}+{{a}_{13}}{{e}^{{{\eta }_{1}}+{{\eta }_{3}}}}+{{a}_{23}}{{e}^{{{\eta }_{2}}+{{\eta }_{3}}}}+{{a}_{123}}{{e}^{{{\eta }_{1}}+{{\eta }_{2}}+{{\eta }_{3}}}}$ to determine ${{a}_{123}}$, which should result in ${{a}_{123}}={{a}_{12}}{{a}_{13}}{{a}_{23}}$ for 3-soliton solutions to exist. The parameter ${{a}_{123}}$ should be in terms of the free parameters ${{a}_{ij}}$ only, because 3-soliton solutions and higher level soliton-solutions should not contain free parameters other than ${{a}_{ij}}$. The existence of 3-soliton solutions confirms the fact that $N$-soliton solutions exist for any order. Based on this, we conclude that the GNNVEs, like the KdV equation, are completely integrable.

{\bf Remark 1.} Eq. (11) can also be written as
\begin{displaymath}
\begin{aligned} {{\left( \frac{\left( aD_{x}^{3}{{D}_{y}}-3a{{a}_{000}}D_{x}^{2}+\left( c-3a{{b}_{000}} \right){{D}_{x}}{{D}_{y}}+dD_{y}^{2} \right)w\cdot w}{{{w}^{2}}} \right)}_{x}} \\
  -b{{\left( \frac{\left( 3{{a}_{000}}D_{y}^{2}+3{{c}_{000}}{{D}_{x}}{{D}_{y}}-{{D}_{x}}D_{y}^{3}+\frac{1}{b}{{D}_{x}}{{D}_{t}} \right)w\cdot w}{{{w}^{2}}} \right)}_{y}}=0. \\
\end{aligned} \eqno(16)
\end{displaymath}
From Eq. (16), we suppose
$$\left( aD_{x}^{3}{{D}_{y}}-3a{{a}_{000}}D_{x}^{2}+\left( c-3a{{b}_{000}} \right){{D}_{x}}{{D}_{y}}+dD_{y}^{2} \right)w\cdot w=0,\eqno           (17)$$
and
$$\left( 3{{a}_{000}}D_{y}^{2}+3{{c}_{000}}{{D}_{x}}{{D}_{y}}-{{D}_{x}}D_{y}^{3}+\frac{1}{b}{{D}_{x}}{{D}_{t}} \right)w\cdot w=0.\eqno             (18)$$
Similarly, to determine the dispersion relation, we set ${{a}_{000}}=0$ and
$$w=1+{{e}^{Px+\Omega y+Kt}}.\eqno                         (19)$$
Substituting Eq. (19) into Eqs. (17) and (18) yields the dispersion relations
$$\Omega =-\frac{P\left( a{{P}^{2}}-3a{{b}_{000}}+c \right)}{d},$$
$$K=\frac{bP\left( a{{P}^{2}}-3a{{b}_{000}}+c \right)\left( {{a}^{2}}{{P}^{6}}+\left( 2ac-6{{a}^{2}}{{b}_{000}} \right){{P}^{4}}+{{\left( 3a{{b}_{000}}-c \right)}^{2}}{{P}^{2}}-3{{c}_{000}}{{d}^{2}} \right)}{{{d}^{3}}}.$$
Accordingly, we get the 1-soliton solution, 2-soliton solution, 3-soliton solution as follows:
$${{w}_{1}}=1+{{e}^{{{\eta }_{1}}}},\ {{w}_{2}}=1+{{e}^{{{\eta }_{1}}}}+{{e}^{{{\eta }_{2}}}}+{{a}_{12}}{{e}^{{{\eta }_{1}}+{{\eta }_{2}}}},$$
$${{w}_{3}}=1+{{e}^{{{\eta }_{1}}}}+{{e}^{{{\eta }_{2}}}}+{{e}^{{{\eta }_{3}}}}+{{a}_{12}}{{e}^{{{\eta }_{1}}+{{\eta }_{2}}}}+{{a}_{13}}{{e}^{{{\eta }_{1}}+{{\eta }_{3}}}}+{{a}_{23}}{{e}^{{{\eta }_{2}}+{{\eta }_{3}}}}+{{a}_{123}}{{e}^{{{\eta }_{1}}+{{\eta }_{2}}+{{\eta }_{3}}}},\eqno      (20)$$
$${{u}_{i}}=-2{{\left( \ln {{w}_{i}} \right)}_{xy}},\ {{v}_{i}}=-2{{\left( \ln {{w}_{i}} \right)}_{xx}}+{{b}_{000}},\ {{\omega }_{i}}=-2{{\left( \ln {{w}_{i}} \right)}_{yy}}+{{c}_{000}},$$
where
$${{\eta }_{i}}={{K}_{i}}t+{{\Omega }_{i}}y+{{P}_{i}}x,\ {{\Omega }_{i}}=-\frac{{{P}_{i}}\left( aP_{i}^{2}-3a{{b}_{000}}+c \right)}{d},$$
$${{K}_{i}}=\frac{b{{P}_{i}}\left( aP_{i}^{2}-3a{{b}_{000}}+c \right)\left( {{a}^{2}}P_{i}^{6}+\left( 2ac-6{{a}^{2}}{{b}_{000}} \right)P_{i}^{4}+{{\left( 3a{{b}_{000}}-c \right)}^{2}}P_{i}^{2}-3{{c}_{000}}{{d}^{2}} \right)}{{{d}^{3}}},$$
$${{a}_{ij}}=\frac{{{\left( {{P}_{i}}-{{P}_{j}} \right)}^{2}}\left( a\left( P_{i}^{2}+{{P}_{i}}{{P}_{j}}+P_{j}^{2}-3{{b}_{000}} \right)+c \right)}{{{\left( {{P}_{i}}+{{P}_{j}} \right)}^{2}}\left( a\left( P_{i}^{2}-{{P}_{i}}{{P}_{j}}+P_{j}^{2}-3{{b}_{000}} \right)+c \right)},\ \left( i,j=1,2,3;i<j \right),\ {{a}_{123}}={{a}_{12}}{{a}_{13}}{{a}_{23}},$$
and ${{P}_{i}}\ \left( {{P}_{i}}\ne 0 \right)\ \left( i=1,2,3 \right),\ {{b}_{000}}$ and ${{c}_{000}}$ are arbitrary constants.

{\bf Remark 2.} Setting ${{a}_{000}}={{b}_{000}}={{c}_{000}}$ in Eq. (11), Eq. (11) can be written as
\begin{displaymath}
\begin{aligned}{{\left( \frac{{{D}_{y}}\left( {{w}_{t}}+a{{w}_{xxx}}+b{{w}_{yyy}}+c{{w}_{x}}+d{{w}_{y}} \right)\cdot w}{{{w}^{2}}} \right)}_{x}}-\frac{3\left( a{{w}_{x}}{{w}_{xxxy}}+b{{w}_{y}}{{w}_{xyyy}}-a{{w}_{xxx}}{{w}_{xy}}-b{{w}_{xy}}{{w}_{yyy}} \right)}{{{w}^{2}}} \\
  -\frac{6\left( a{{w}_{x}}{{w}_{xx}}{{w}_{xy}}+b{{w}_{y}}{{w}_{yy}}{{w}_{xy}}-aw_{x}^{2}{{w}_{xxy}}-bw_{y}^{2}{{w}_{xyy}} \right)}{{{w}^{3}}}=0. \\
\end{aligned}
\end{displaymath}
We suppose that
$${{w}_{t}}+a{{w}_{xxx}}+b{{w}_{yyy}}+c{{w}_{x}}+d{{w}_{y}}=0,\eqno                     (21a)$$
$$a{{w}_{x}}{{w}_{xxxy}}+b{{w}_{y}}{{w}_{xyyy}}-a{{w}_{xxx}}{{w}_{xy}}-b{{w}_{xy}}{{w}_{yyy}}=0,\eqno                   (21b)$$
$$a{{w}_{x}}{{w}_{xx}}{{w}_{xy}}+b{{w}_{y}}{{w}_{yy}}{{w}_{xy}}-aw_{x}^{2}{{w}_{xxy}}-bw_{y}^{2}{{w}_{xyy}}=0.\eqno              (21c)$$
Suppose that the solution of Eqs. (21) is the form
$$w=A+Bf\left( \xi  \right){{e}^{\eta }},\eqno                           (22)$$
where $\xi =Kx+Ly+\Omega t+{{\xi }_{0}}$, $\eta =kx+ly+\varpi t+{{\eta }_{0}}$, and $A,B,K,L,\Omega ,{{\xi }_{0}},k,l,\varpi ,{{\eta }_{0}}$ are constants to be determined later.
Substituting Eq. (22) into Eqs. (21) yields a  set of algebraic equations for $A,B,K,L,\Omega ,{{\xi }_{0}},k,l,\varpi ,{{\eta }_{0}}$. Accordingly, $w,u,v$ and $\omega $ can be determined [29].

\section{Exact three wave solutions of the GNNVEs}
In this section, the three-wave method [33] is employed to seek the exact solutions of the GNNVEs.
With regard to Eqs. (12) and (13), using three-wave method, we are going to seek the solution of the form
$$w={{e}^{-{{\xi }_{1}}}}+{{d}_{1}}\cos {{\xi }_{2}}+{{d}_{2}}\cosh {{\xi }_{3}}+{{d}_{3}}{{e}^{{{\xi }_{1}}}},\eqno                  (23)$$
where ${{\xi }_{i}}={{\alpha }_{i}}x+{{\beta }_{i}}y+{{\gamma }_{i}}t\ \left( i=1,2,3 \right)$, and ${{\alpha }_{i}},{{\beta }_{i}},\ {{\gamma }_{i}}$ and ${{d}_{i}}\ \left( i=1,2,3 \right)$are constants to determined later.\\
Substituting Eq. (23) into Eqs. (12) and (13), and equating all coefficients of ${{e}^{\pm {{\xi }_{1}}}}\sinh {{\xi }_{3}}$,\ ${{e}^{\pm {{\xi }_{1}}}}\cosh {{\xi }_{3}}$,\ ${{e}^{\pm {{\xi }_{1}}}}\sin {{\xi }_{3}}$,\ ${{e}^{\pm {{\xi }_{1}}}}\cos {{\xi }_{3}}$,\ $\sinh {{\xi }_{3}}\sin {{\xi }_{2}}$,\ $\cosh {{\xi }_{3}}\cos {{\xi }_{2}}$ and constant term to zero, we obtain a set of algebraic equations for ${{\alpha }_{i}},{{\beta }_{i}},{{\gamma }_{i}},{{d}_{i}}\ \left( i=1,2,3 \right), {{a}_{000}},{{b}_{000}}$ and ${{c}_{000}}$.\\
From Eq. (12), we have\\
$a\alpha _{1}^{3}{{\beta }_{3}}{{d}_{2}}{{d}_{3}}+3a\alpha _{1}^{2}{{\alpha }_{3}}{{\beta }_{1}}{{d}_{2}}{{d}_{3}}+3a{{\alpha }_{1}}\alpha _{3}^{2}{{\beta }_{3}}{{d}_{2}}{{d}_{3}}+a\alpha _{3}^{3}{{\beta }_{1}}{{d}_{2}}{{d}_{3}}-6a{{a}_{000}}{{\alpha }_{1}}{{\alpha }_{3}}{{d}_{2}}{{d}_{3}}+2d{{\beta }_{1}}{{\beta }_{3}}{{d}_{2}}{{d}_{3}}
  +{{\beta }_{3}}{{\gamma }_{1}}{{d}_{2}}{{d}_{3}}-3a{{b}_{000}}{{\alpha }_{1}}{{\beta }_{3}}{{d}_{2}}{{d}_{3}}-3a{{b}_{000}}{{\alpha }_{3}}{{\beta }_{1}}{{d}_{2}}{{d}_{3}}+c{{\alpha }_{1}}{{\beta }_{3}}{{d}_{2}}{{d}_{3}}+c{{\alpha }_{3}}{{\beta }_{1}}{{d}_{2}}{{d}_{3}}+{{\beta }_{1}}{{\gamma }_{3}}{{d}_{2}}{{d}_{3}}=0,
$\\
$3a{{b}_{000}}{{\alpha }_{2}}{{\beta }_{3}}{{d}_{1}}{{d}_{2}}+3a\alpha _{2}^{2}{{\alpha }_{3}}{{\beta }_{2}}{{d}_{1}}{{d}_{2}}-3a{{\alpha }_{2}}\alpha _{3}^{2}{{\beta }_{3}}{{d}_{1}}{{d}_{2}}-a\alpha _{3}^{3}{{\beta }_{2}}{{d}_{1}}{{d}_{2}}+6a{{a}_{000}}{{\alpha }_{2}}{{\alpha }_{3}}{{d}_{1}}{{d}_{2}}-{{\beta }_{2}}{{\gamma }_{3}}{{d}_{1}}{{d}_{2}}
  +a\alpha _{2}^{3}{{\beta }_{3}}{{d}_{1}}{{d}_{2}}+3a{{b}_{000}}{{\alpha }_{3}}{{\beta }_{2}}{{d}_{1}}{{d}_{2}}-c{{\alpha }_{2}}{{\beta }_{3}}{{d}_{1}}{{d}_{2}}-c{{\alpha }_{3}}{{\beta }_{2}}{{d}_{1}}{{d}_{2}}-2d{{\beta }_{2}}{{\beta }_{3}}{{d}_{1}}{{d}_{2}}-{{\beta }_{3}}{{\gamma }_{2}}{{d}_{1}}{{d}_{2}}=0,$\\
$3a{{b}_{000}}{{\alpha }_{1}}{{\beta }_{3}}{{d}_{2}}-a\alpha _{1}^{3}{{\beta }_{3}}{{d}_{2}}-3a\alpha _{1}^{2}{{\alpha }_{3}}{{\beta }_{1}}{{d}_{2}}-3a{{\alpha }_{1}}\alpha _{3}^{2}{{\beta }_{3}}{{d}_{2}}-a\alpha _{3}^{3}{{\beta }_{1}}{{d}_{2}}+6a{{a}_{000}}{{\alpha }_{1}}{{\alpha }_{3}}{{d}_{2}}
 +3a{{b}_{000}}{{\alpha }_{3}}{{\beta }_{1}}{{d}_{2}}-c{{\alpha }_{1}}{{\beta }_{3}}{{d}_{2}}-c{{\alpha }_{3}}{{\beta }_{1}}{{d}_{2}}-2d{{\beta }_{1}}{{\beta }_{3}}{{d}_{2}}-{{\beta }_{1}}{{\gamma }_{3}}{{d}_{2}}-{{\beta }_{3}}{{\gamma }_{1}}{{d}_{2}}=0,$\\
$a\alpha _{1}^{3}{{\beta }_{2}}{{d}_{1}}+3a\alpha _{1}^{2}{{\alpha }_{2}}{{\beta }_{1}}{{d}_{1}}-3a{{\alpha }_{1}}\alpha _{2}^{2}{{\beta }_{2}}{{d}_{1}}-a\alpha _{2}^{3}{{\beta }_{1}}{{d}_{1}}-6a{{a}_{000}}{{\alpha }_{1}}{{\alpha }_{2}}{{d}_{1}}-3a{{b}_{000}}{{\alpha }_{1}}{{\beta }_{2}}{{d}_{1}}
 -3a{{b}_{000}}{{\alpha }_{2}}{{\beta }_{1}}{{d}_{1}}+c{{\alpha }_{1}}{{\beta }_{2}}{{d}_{1}}+c{{\alpha }_{2}}{{\beta }_{1}}{{d}_{1}}+2d{{\beta }_{1}}{{\beta }_{2}}{{d}_{1}}+{{\beta }_{1}}{{\gamma }_{2}}{{d}_{1}}+{{\beta }_{2}}{{\gamma }_{1}}{{d}_{1}}=0,$\\
$3a{{b}_{000}}{{\alpha }_{2}}{{\beta }_{1}}{{d}_{1}}{{d}_{3}}-a\alpha _{1}^{3}{{\beta }_{2}}{{d}_{1}}{{d}_{3}}-3a\alpha _{1}^{2}{{\alpha }_{2}}{{\beta }_{1}}{{d}_{1}}{{d}_{3}}+a\alpha _{2}^{3}{{\beta }_{1}}{{d}_{1}}{{d}_{3}}+6a{{a}_{000}}{{\alpha }_{1}}{{\alpha }_{2}}{{d}_{1}}{{d}_{3}}-{{\beta }_{1}}{{\gamma }_{2}}{{d}_{1}}{{d}_{3}}
 +3a{{\alpha }_{1}}\alpha _{2}^{2}{{\beta }_{2}}{{d}_{1}}{{d}_{3}}-c{{\alpha }_{1}}{{\beta }_{2}}{{d}_{1}}{{d}_{3}}-c{{\alpha }_{2}}{{\beta }_{1}}{{d}_{1}}{{d}_{3}}-2d{{\beta }_{1}}{{\beta }_{2}}{{d}_{1}}{{d}_{3}}+3a{{b}_{000}}{{\alpha }_{1}}{{\beta }_{2}}{{d}_{1}}{{d}_{3}}-{{\beta }_{2}}{{\gamma }_{1}}{{d}_{1}}{{d}_{3}}=0,
$\\
$3a{{a}_{000}}\alpha _{1}^{2}{{d}_{2}}{{d}_{3}}-a\alpha _{1}^{3}{{\beta }_{1}}{{d}_{2}}{{d}_{3}}-3a\alpha _{1}^{2}{{\alpha }_{3}}{{\beta }_{3}}{{d}_{2}}{{d}_{3}}-3a{{\alpha }_{1}}\alpha _{3}^{2}{{\beta }_{1}}{{d}_{2}}{{d}_{3}}-a\alpha _{3}^{3}{{\beta }_{3}}{{d}_{2}}{{d}_{3}}
 +3a{{a}_{000}}\alpha _{3}^{2}{{d}_{2}}{{d}_{3}}+$\\
$3a{{b}_{000}}{{\alpha }_{1}}{{\beta }_{1}}{{d}_{2}}{{d}_{3}}+3a{{b}_{000}}{{\alpha }_{3}}{{\beta }_{3}}{{d}_{2}}{{d}_{3}}-c{{\alpha }_{1}}{{\beta }_{1}}{{d}_{2}}{{d}_{3}}-c{{\alpha }_{3}}{{\beta }_{3}}{{d}_{2}}{{d}_{3}}
  -d\beta _{1}^{2}{{d}_{2}}{{d}_{3}}-d\beta _{3}^{2}{{d}_{2}}{{d}_{3}}-\quad \quad \quad \  (24)$\\ ${{\beta }_{1}}{{\gamma }_{1}}{{d}_{2}}{{d}_{3}}-{{\beta }_{3}}{{\gamma }_{3}}{{d}_{2}}{{d}_{3}}=0,
$\\
$3a\alpha_{1}^{2}{{\alpha}_{2}}{{\beta}_{2}}{{d}_{1}}{{d}_{3}}-a\alpha_{1}^{3}{{\beta}_{1}}{{d}_{1}}{{d}_{3}}+3a{{\alpha}_{1}}\alpha_{2}^{2}{{\beta}_{1}}{{d}_{1}}{{d}_{3}}-a\alpha _{2}^{3}{{\beta}_{2}}{{d}_{1}}{{d}_{3}}+3a{{a}_{000}}\alpha_{1}^{2}{{d}_{1}}{{d}_{3}}-d\beta_{1}^{2}{{d}_{1}}{{d}_{3}}-$\\ $
3{{a}_{000}}a\alpha_{2}^{2}{{d}_{1}}{{d}_{3}}+3a{{b}_{000}}{{\alpha}_{1}}{{\beta}_{1}}{{d}_{1}}{{d}_{3}}-3a{{b}_{000}}{{\alpha}_{2}}{{\beta}_{2}}{{d}_{1}}{{d}_{3}}-c{{\alpha}_{1}}{{\beta }_{1}}{{d}_{1}}{{d}_{3}}+c{{\alpha }_{2}}{{\beta}_{2}}{{d}_{1}}{{d}_{3}}
 +d\beta_{2}^{2}{{d}_{1}}{{d}_{3}}-{{\beta}_{1}}{{\gamma}_{1}}{{d}_{1}}{{d}_{3}}+{{\beta}_{2}}{{\gamma}_{2}}{{d}_{1}}{{d}_{3}}=0,
$\\
$3a\alpha _{1}^{2}{{\alpha }_{2}}{{\beta }_{2}}{{d}_{1}}-3a{{b}_{000}}{{\alpha }_{2}}{{\beta }_{2}}{{d}_{1}}+3a{{\alpha }_{1}}\alpha _{2}^{2}{{\beta }_{1}}{{d}_{1}}+3a{{a}_{000}}\alpha _{1}^{2}{{d}_{1}}-3a{{a}_{000}}\alpha _{2}^{2}{{d}_{1}}+3a{{\alpha }_{1}}{{b}_{000}}{{\beta }_{1}}{{d}_{1}}
 -a\alpha _{1}^{3}{{\beta }_{1}}{{d}_{1}}-a\alpha _{2}^{3}{{\beta }_{2}}{{d}_{1}}-c{{\alpha }_{1}}{{\beta }_{1}}{{d}_{1}}+c{{\alpha }_{2}}{{\beta }_{2}}{{d}_{1}}-d\beta _{1}^{2}{{d}_{1}}+d\beta _{2}^{2}{{d}_{1}}-{{\beta }_{1}}{{\gamma }_{1}}{{d}_{1}}+{{\beta }_{2}}{{\gamma }_{2}}{{d}_{1}}=0,
$\\
$3a{{a}_{000}}\alpha _{1}^{2}{{d}_{2}}-a\alpha _{1}^{3}{{\beta }_{1}}{{d}_{2}}-3a\alpha _{1}^{2}{{\alpha }_{3}}{{\beta }_{3}}{{d}_{2}}-3a{{\alpha }_{1}}\alpha _{3}^{2}{{\beta }_{1}}{{d}_{2}}-a\alpha _{3}^{3}{{\beta }_{3}}{{d}_{2}}+3a{{a}_{000}}\alpha _{3}^{2}{{d}_{2}}-{{\beta }_{1}}{{\gamma }_{1}}{{d}_{2}}
 +3a{{b}_{000}}{{\alpha }_{1}}{{\beta }_{1}}{{d}_{2}}+3a{{b}_{000}}{{\alpha }_{3}}{{\beta }_{3}}{{d}_{2}}-c{{\alpha }_{1}}{{\beta }_{1}}{{d}_{2}}-c{{\alpha }_{3}}{{\beta }_{3}}{{d}_{2}}-d\beta _{1}^{2}{{d}_{2}}-d\beta _{3}^{2}{{d}_{2}}-{{\beta }_{3}}{{\gamma }_{3}}{{d}_{2}}=0,
$\\
$3a\alpha _{2}^{2}{{\alpha }_{3}}{{\beta }_{3}}{{d}_{1}}{{d}_{2}}-a\alpha _{2}^{3}{{\beta }_{2}}{{d}_{1}}{{d}_{2}}+3a{{\alpha }_{2}}\alpha _{3}^{2}{{\beta }_{2}}{{d}_{1}}{{d}_{2}}-a\alpha _{3}^{3}{{\beta }_{3}}{{d}_{1}}{{d}_{2}}-3a{{a}_{000}}\alpha _{2}^{2}{{d}_{1}}{{d}_{2}}
 +3a{{a}_{000}}\alpha _{3}^{2}{{d}_{1}}{{d}_{2}}-3a{{b}_{000}}{{\alpha }_{2}}{{\beta }_{2}}{{d}_{1}}{{d}_{2}}+3a{{b}_{000}}{{\alpha }_{3}}{{\beta }_{3}}{{d}_{1}}{{d}_{2}}+c{{\alpha }_{2}}{{\beta }_{2}}{{d}_{1}}{{d}_{2}}-c{{\alpha }_{3}}{{\beta }_{3}}{{d}_{1}}{{d}_{2}}
 +d\beta _{2}^{2}{{d}_{1}}{{d}_{2}}-d\beta _{3}^{2}{{d}_{1}}{{d}_{2}}+{{\beta }_{2}}{{\gamma }_{2}}{{d}_{1}}{{d}_{2}}-{{\beta }_{3}}{{\gamma }_{3}}{{d}_{1}}{{d}_{2}}=0, $\\
$3a{{a}_{000}}\alpha _{3}^{2}d_{2}^{2}-4a\alpha _{2}^{3}{{\beta }_{2}}d_{1}^{2}-4a\alpha _{3}^{3}{{\beta }_{3}}d_{2}^{2}-3a{{a}_{000}}\alpha _{2}^{2}d_{1}^{2}+3a{{b}_{000}}{{\alpha }_{3}}{{\beta }_{3}}d_{2}^{2}-16a\alpha _{1}^{3}{{\beta }_{1}}{{d}_{3}}
 -4{{\beta }_{1}}{{\gamma }_{1}}{{d}_{3}}-$\\ $3a{{b}_{000}}{{\alpha }_{2}}{{\beta }_{2}}d_{1}^{2}+12a{{a}_{000}}\alpha _{1}^{2}{{d}_{3}}+12a{{b}_{000}}{{\alpha }_{1}}{{\beta }_{1}}{{d}_{3}}+c{{\alpha }_{2}}{{\beta }_{2}}d_{1}^{2}-c{{\alpha }_{3}}{{\beta }_{3}}d_{2}^{2}+d\beta _{2}^{2}d_{1}^{2}
 -d\beta _{3}^{2}d_{2}^{2}-4c{{\alpha }_{1}}{{\beta }_{1}}{{d}_{3}}-4d\beta _{1}^{2}{{d}_{3}}+{{\beta }_{2}}{{\gamma }_{2}}d_{1}^{2}-{{\beta }_{3}}{{\gamma }_{3}}d_{2}^{2}=0.$\\
And from Eq. (13), we have\\
$3{{\alpha }_{2}}\beta _{2}^{2}{{\beta }_{3}}{{d}_{1}}{{d}_{2}}-{{\alpha }_{2}}\beta _{3}^{3}{{d}_{1}}{{d}_{2}}+{{\alpha }_{3}}\beta _{2}^{3}{{d}_{1}}{{d}_{2}}-3{{\alpha }_{3}}{{\beta }_{2}}\beta _{3}^{2}{{d}_{1}}{{d}_{2}}+6{{a}_{000}}{{\beta }_{2}}{{\beta }_{3}}{{d}_{1}}{{d}_{2}}
 +3{{c}_{000}}{{\alpha }_{2}}{{\beta }_{3}}{{d}_{1}}{{d}_{2}}+3{{c}_{000}}{{\alpha }_{3}}{{\beta }_{2}}{{d}_{1}}{{d}_{2}}=0,$\\
$6{{a}_{000}}{{\beta }_{1}}{{\beta }_{3}}{{d}_{2}}-3{{\alpha }_{1}}\beta _{1}^{2}{{\beta }_{3}}{{d}_{2}}-{{\alpha }_{1}}\beta _{3}^{3}{{d}_{2}}-{{\alpha }_{3}}\beta _{1}^{3}{{d}_{2}}-3{{\alpha }_{3}}{{\beta }_{1}}\beta _{3}^{2}{{d}_{2}}+3{{c}_{000}}{{\alpha }_{1}}{{\beta }_{3}}{{d}_{2}}+3{{c}_{000}}{{\alpha }_{3}}{{\beta }_{1}}{{d}_{2}}=0,$\\
$3{{\alpha }_{1}}\beta _{1}^{2}{{\beta }_{3}}{{d}_{2}}{{d}_{3}}+{{\alpha }_{1}}\beta _{3}^{3}{{d}_{2}}{{d}_{3}}+{{\alpha }_{3}}\beta _{1}^{3}{{d}_{2}}{{d}_{3}}+3{{\alpha }_{3}}{{\beta }_{1}}\beta _{3}^{2}{{d}_{2}}{{d}_{3}}-6{{a}_{000}}{{\beta }_{1}}{{\beta }_{3}}{{d}_{2}}{{d}_{3}}
 -3{{c}_{000}}{{\alpha }_{1}}{{\beta }_{3}}{{d}_{2}}{{d}_{3}}-3{{c}_{000}}{{\alpha }_{3}}{{\beta }_{1}}{{d}_{2}}{{d}_{3}}=0,
$\\
${{\alpha }_{1}}\beta _{2}^{3}{{d}_{1}}{{d}_{3}}-3{{\alpha }_{1}}\beta _{1}^{2}{{\beta }_{2}}{{d}_{1}}{{d}_{3}}-{{\alpha }_{2}}\beta _{1}^{3}{{d}_{1}}{{d}_{3}}+3{{\alpha }_{2}}{{\beta }_{1}}\beta _{2}^{2}{{d}_{1}}{{d}_{3}}+6{{a}_{000}}{{\beta }_{1}}{{\beta }_{2}}{{d}_{1}}{{d}_{3}}
 +3{{c}_{000}}{{\alpha }_{1}}{{\beta }_{2}}{{d}_{1}}{{d}_{3}}+3{{c}_{000}}{{\alpha }_{2}}{{\beta }_{1}}{{d}_{1}}{{d}_{3}}=0,
$\\
$3{{\alpha }_{1}}\beta _{1}^{2}{{\beta }_{2}}{{d}_{1}}-{{\alpha }_{1}}\beta _{2}^{3}{{d}_{1}}+{{\alpha }_{2}}\beta _{1}^{3}{{d}_{1}}-3{{\alpha }_{2}}{{\beta }_{1}}\beta _{2}^{2}{{d}_{1}}-6{{a}_{000}}{{\beta }_{1}}{{\beta }_{2}}{{d}_{1}}-3{{c}_{000}}{{\alpha }_{1}}{{\beta }_{2}}{{d}_{1}}-3{{c}_{000}}{{\alpha }_{2}}{{\beta }_{1}}{{d}_{1}}=0,$\\
$3{{\alpha }_{1}}{{\beta }_{1}}\beta _{2}^{2}{{d}_{1}}{{d}_{3}}-{{\alpha }_{1}}\beta _{1}^{3}{{d}_{1}}{{d}_{3}}+3{{\alpha }_{2}}\beta _{1}^{2}{{\beta }_{2}}{{d}_{1}}{{d}_{3}}-{{\alpha }_{2}}\beta _{2}^{3}{{d}_{1}}{{d}_{3}}+3{{a}_{000}}\beta _{1}^{2}{{d}_{1}}{{d}_{3}}
 -3{{a}_{000}}\beta _{2}^{2}{{d}_{1}}{{d}_{3}}+\ \quad \quad \ (25)$\\ $3{{c}_{000}}{{\alpha }_{1}}{{\beta }_{1}}{{d}_{1}}{{d}_{3}}-3{{c}_{000}}{{\alpha }_{2}}{{\beta }_{2}}{{d}_{1}}{{d}_{3}}=0, $\\
$3{{a}_{000}}\beta _{3}^{2}{{d}_{2}}{{d}_{3}}-{{\alpha }_{1}}\beta _{1}^{3}{{d}_{2}}{{d}_{3}}-3{{\alpha }_{1}}{{\beta }_{1}}\beta _{3}^{2}{{d}_{2}}{{d}_{3}}-3{{\alpha }_{3}}\beta _{1}^{2}{{\beta }_{3}}{{d}_{2}}{{d}_{3}}-{{\alpha }_{3}}\beta _{3}^{3}{{d}_{2}}{{d}_{3}}+3{{a}_{000}}\beta _{1}^{2}{{d}_{2}}{{d}_{3}}
 +3{{c}_{000}}{{\alpha }_{1}}{{\beta }_{1}}{{d}_{2}}{{d}_{3}}+3{{c}_{000}}{{\alpha }_{3}}{{\beta }_{3}}{{d}_{2}}{{d}_{3}}=0,
$\\
$3{{\alpha}_{1}}{{\beta}_{1}}\beta_{2}^{2}{{d}_{1}}-{{\alpha}_{1}}\beta_{1}^{3}{{d}_{1}}+3{{\alpha}_{2}}\beta_{1}^{2}{{\beta}_{2}}{{d}_{1}}-{{\alpha}_{2}}\beta _{2}^{3}{{d}_{1}}+3{{a}_{000}}\beta _{1}^{2}{{d}_{1}}-3{{a}_{000}}\beta_{2}^{2}{{d}_{1}}+3{{c}_{000}}{{\alpha}_{1}}{{\beta}_{1}}{{d}_{1}}-$\\ $3{{c}_{000}}{{\alpha}_{2}}{{\beta}_{2}}{{d}_{1}}=0,
$\\
$3{{a}_{000}}\beta _{1}^{2}{{d}_{2}}-{{\alpha }_{1}}\beta _{1}^{3}{{d}_{2}}-3{{\alpha }_{1}}{{\beta }_{1}}\beta _{3}^{2}{{d}_{2}}-3{{\alpha }_{3}}\beta _{1}^{2}{{\beta }_{3}}{{d}_{2}}-{{\alpha }_{3}}\beta _{3}^{3}{{d}_{2}}
 +3{{a}_{000}}\beta _{3}^{2}{{d}_{2}}+3{{c}_{000}}{{\alpha }_{1}}{{\beta }_{1}}{{d}_{2}}+$\\ $3{{c}_{000}}{{\alpha }_{3}}{{\beta }_{3}}{{d}_{2}}=0,
$\\
$3{{\alpha }_{2}}{{\beta }_{2}}\beta _{3}^{2}{{d}_{1}}{{d}_{2}}-{{\alpha }_{2}}\beta _{2}^{3}{{d}_{1}}{{d}_{2}}+3{{\alpha }_{3}}\beta _{2}^{2}{{\beta }_{3}}{{d}_{1}}{{d}_{2}}-{{\alpha }_{3}}\beta _{3}^{3}{{d}_{1}}{{d}_{2}}-3{{a}_{000}}\beta _{2}^{2}{{d}_{1}}{{d}_{2}}
 +3{{a}_{000}}\beta _{3}^{2}{{d}_{1}}{{d}_{2}}-$\\ $ 3{{c}_{000}}{{\alpha }_{2}}{{\beta }_{2}}{{d}_{1}}{{d}_{2}}+3{{c}_{000}}{{\alpha }_{3}}{{\beta }_{3}}{{d}_{1}}{{d}_{2}}=0, $\\
$3{{a}_{000}}\beta _{3}^{2}d_{2}^{2}-4{{\alpha }_{2}}\beta _{2}^{3}d_{1}^{2}-4{{\alpha }_{3}}\beta _{3}^{3}d_{2}^{2}-3{{a}_{000}}\beta _{2}^{2}d_{1}^{2}-16{{\alpha }_{1}}\beta _{1}^{3}{{d}_{3}}-3{{c}_{000}}{{\alpha }_{2}}{{\beta }_{2}}d_{1}^{2}
 +3{{c}_{000}}{{\alpha }_{3}}{{\beta }_{3}}d_{2}^{2}+12{{a}_{000}}\beta _{1}^{2}{{d}_{3}}+12{{c}_{000}}{{\alpha }_{1}}{{\beta }_{1}}{{d}_{3}}=0.
$

Denote that $I=\sqrt{-1}$ and $\varepsilon =\pm 1$. Solving the systems of (24) and (25) with the aid of Maple, we obtain the following cases:

{\bf Case 1.} ${{d}_{1}}=0$, ${{d}_{2}}=0$, ${{d}_{3}}={{d}_{3}}$.\\
(1) ${{a}_{000}}=\frac{{{\alpha }_{1}}\left( 4\beta _{1}^{2}-3{{c}_{000}} \right)}{3{{\beta }_{1}}}$, ${{b}_{000}}={{b}_{000}}$, ${{c}_{000}}={{c}_{000}}$, ${{\alpha }_{1}}={{\alpha }_{1}}$, ${{\alpha }_{2}}={{\alpha }_{2}}$, ${{\alpha }_{3}}={{\alpha }_{3}}$, ${{\beta }_{1}}={{\beta }_{1}}$, ${{\beta }_{2}}={{\beta }_{2}}$, ${{\beta }_{3}}={{\beta }_{3}}$, ${{\gamma }_{1}}=-\frac{\left( 3a{{c}_{000}}\alpha _{1}^{3}-3a{{b}_{000}}{{\alpha }_{1}}\beta _{1}^{2}+c{{\alpha }_{1}}\beta _{1}^{2}+d\beta _{1}^{3} \right)}{\beta _{1}^{2}}$, ${{\gamma }_{2}}={{\gamma }_{2}}$, ${{\gamma }_{3}}={{\gamma }_{3}}$.\\
Accordingly, we get ${{w}_{1}}={{e}^{-{{\xi }_{1}}}}+{{d}_{3}}{{e}^{{{\xi }_{1}}}}$, where
${{\xi }_{1}}={{\alpha }_{1}}x+{{\beta }_{1}}y-\frac{\left( 3a{{c}_{000}}\alpha _{1}^{3}-3a{{b}_{000}}{{\alpha }_{1}}\beta _{1}^{2}+c{{\alpha }_{1}}\beta _{1}^{2}+d\beta _{1}^{3} \right)t}{\beta _{1}^{2}}$.

{\bf Case 2.} ${{d}_{1}}=0$, ${{d}_{2}}={{d}_{2}}$, ${{d}_{3}}=0$.\\
(1) ${{a}_{000}}=\frac{{{\alpha }_{1}}\left( \beta _{1}^{2}-3{{c}_{000}} \right)}{3{{\beta }_{1}}}$, ${{b}_{000}}={{b}_{000}}$, ${{c}_{000}}={{c}_{000}}$, ${{\alpha }_{1}}={{\alpha }_{1}}$, ${{\alpha }_{2}}={{\alpha }_{2}}$, ${{\alpha }_{3}}=0$, ${{\beta }_{1}}={{\beta }_{1}}$, ${{\beta }_{2}}={{\beta }_{2}}$, ${{\beta }_{3}}=0$, ${{\gamma }_{1}}=-\frac{3a{{c}_{000}}\alpha _{1}^{3}-3a{{b}_{000}}{{\alpha }_{1}}\beta _{1}^{2}+c{{\alpha }_{1}}\beta _{1}^{2}+d\beta _{1}^{3}}{\beta _{1}^{2}}$, ${{\gamma }_{2}}={{\gamma }_{2}}$, ${{\gamma }_{3}}=0$.\\
(2) ${{a}_{000}}=0$, ${{b}_{000}}={{b}_{000}}$, ${{c}_{000}}=\frac{\beta _{1}^{2}}{3}$, ${{\alpha }_{1}}={{\alpha }_{1}}$, ${{\alpha }_{2}}={{\alpha }_{2}}$, ${{\alpha }_{3}}={{\alpha }_{3}}$, ${{\beta }_{1}}={{\beta }_{1}}$, ${{\beta }_{2}}={{\beta }_{2}}$, ${{\beta }_{3}}=0$, ${{\gamma }_{1}}=3a{{b}_{000}}{{\alpha }_{1}}-a\alpha _{1}^{3}-3a{{\alpha }_{1}}\alpha _{3}^{2}-c{{\alpha }_{1}}-d{{\beta }_{1}}$, ${{\gamma }_{2}}={{\gamma }_{2}}$,
${{\gamma }_{3}}=3a{{b}_{000}}{{\alpha }_{3}}-3a\alpha _{1}^{2}{{\alpha }_{3}}-a\alpha _{3}^{3}-c{{\alpha }_{3}}$.\\
Accordingly, we get (1) ${{w}_{2}}={{e}^{-{{\xi }_{1}}}}+{{d}_{2}}$, (2) ${{w}_{2}}={{e}^{-{{\xi }_{1}}}}+{{d}_{2}}\cosh {{\xi }_{3}}$, where\\
(1) ${{\xi }_{1}}={{\alpha }_{1}}x+{{\beta }_{1}}y-\left( \frac{3a{{c}_{000}}\alpha _{1}^{3}-3a{{b}_{000}}{{\alpha }_{1}}\beta _{1}^{2}+c{{\alpha }_{1}}\beta _{1}^{2}+d\beta _{1}^{3}}{\beta _{1}^{2}} \right)t$.\\
(2) ${{\xi }_{1}}={{\alpha }_{1}}x+{{\beta }_{1}}y+\left( 3a{{b}_{000}}{{\alpha }_{1}}-a\alpha _{1}^{3}-3a{{\alpha }_{1}}\alpha _{3}^{2}-c{{\alpha }_{1}}-d{{\beta }_{1}} \right)t$,\\
${{\xi }_{3}}={{\alpha }_{3}}x+\left( 3a{{b}_{000}}{{\alpha }_{3}}-3a\alpha _{1}^{2}{{\alpha }_{3}}-a\alpha _{3}^{3}-c{{\alpha }_{3}} \right)t$.

{\bf Case 3.} ${{d}_{1}}={{d}_{1}}$, ${{d}_{2}}=0$, ${{d}_{3}}=0$.\\
(1) ${{a}_{000}}=0$, ${{b}_{000}}={{b}_{000}}$, ${{c}_{000}}=\frac{\beta _{1}^{2}}{3}$, ${{\alpha }_{1}}={{\alpha }_{1}}$, ${{\alpha }_{2}}={{\alpha }_{2}}$, ${{\alpha }_{3}}={{\alpha }_{3}}$, ${{\beta }_{1}}={{\beta }_{1}}$, ${{\beta }_{2}}=0$, ${{\beta }_{3}}={{\beta }_{3}}$, ${{\gamma }_{1}}=3a{{\alpha }_{1}}\alpha _{2}^{2}-a\alpha _{1}^{3}+3a{{b}_{000}}{{\alpha }_{1}}-c{{\alpha }_{1}}-d{{\beta }_{1}}$, ${{\gamma }_{3}}={{\gamma }_{3}}$,
${{\gamma }_{2}}=3a{{b}_{000}}{{\alpha }_{2}}-3a\alpha _{1}^{2}{{\alpha }_{2}}+a\alpha _{2}^{3}-c{{\alpha }_{2}}$.\\
(2) ${{a}_{000}}=0$, ${{b}_{000}}={{b}_{000}}$, ${{c}_{000}}=-\frac{4\beta _{2}^{2}}{3}$, ${{\alpha }_{1}}={{\alpha }_{1}}$, ${{\alpha }_{2}}=0$, ${{\alpha }_{3}}={{\alpha }_{3}}$, ${{\beta }_{1}}=\varepsilon I{{\beta }_{2}}$, ${{\beta }_{2}}={{\beta }_{2}}$, ${{\beta }_{3}}={{\beta }_{3}}$, ${{\gamma }_{1}}=3a{{b}_{000}}{{\alpha }_{1}}-a\alpha _{1}^{3}-\varepsilon Id{{\beta }_{2}}-c{{\alpha }_{1}}$, ${{\gamma }_{2}}=-d{{\beta }_{2}}$, ${{\gamma }_{3}}={{\gamma }_{3}}$.\\
Accordingly, we get ${{w}_{3}}={{e}^{-{{\xi }_{1}}}}+{{d}_{1}}\cos {{\xi }_{2}}$, where\\
(1) ${{\xi }_{1}}={{\alpha }_{1}}x+{{\beta }_{1}}y+\left( 3a{{\alpha }_{1}}\alpha _{2}^{2}-a\alpha _{1}^{3}+3a{{b}_{000}}{{\alpha }_{1}}-c{{\alpha }_{1}}-d{{\beta }_{1}} \right)t$,\\
${{\xi }_{2}}={{\alpha }_{2}}x+\left( 3a{{b}_{000}}{{\alpha }_{2}}-3a\alpha _{1}^{2}{{\alpha }_{2}}+a\alpha _{2}^{3}-c{{\alpha }_{2}} \right)t$,\\
(2) ${{\xi }_{1}}={{\alpha }_{1}}x+\varepsilon I{{\beta }_{2}}y+\left( 3a{{b}_{000}}{{\alpha }_{1}}-a\alpha _{1}^{3}-\varepsilon Id{{\beta }_{2}}-c{{\alpha }_{1}} \right)t$, ${{\xi }_{2}}={{\beta }_{2}}y-d{{\beta }_{2}}t$.

{\bf Case 4.} ${{d}_{1}}=0$, ${{d}_{2}}={{d}_{2}}$, ${{d}_{3}}={{d}_{3}}$.\\
(1) ${{a}_{000}}=0$, ${{b}_{000}}={{b}_{000}}$, ${{c}_{000}}=\frac{\beta _{3}^{2}}{3}$, ${{\alpha }_{1}}={{\alpha }_{1}}$, ${{\alpha }_{2}}={{\alpha }_{2}}$, ${{\alpha }_{3}}=0$, ${{\beta }_{1}}=0$, ${{\beta }_{2}}={{\beta }_{2}}$, ${{\beta }_{3}}={{\beta }_{3}}$, ${{\gamma }_{1}}=-{{\alpha }_{1}}\left( a\alpha _{1}^{2}-3a{{b}_{000}}+c \right)$, ${{\gamma }_{2}}={{\gamma }_{2}}$, ${{\gamma }_{3}}=-{{\beta }_{3}}d$.\\
(2) ${{a}_{000}}=0$, ${{b}_{000}}={{b}_{000}}$, ${{c}_{000}}=\frac{4\beta _{1}^{2}}{3}$, ${{\alpha }_{1}}={{\alpha }_{1}}$, ${{\alpha }_{2}}={{\alpha }_{2}}$, ${{\alpha }_{3}}={{\alpha }_{3}}$, ${{\beta }_{2}}={{\beta }_{2}}$,
${{\beta}_{1}}={{\beta}_{3}}=-\frac{4ad_{2}^{2}\alpha_{3}^{3}+12a{{d}_{3}}\alpha _{1}^{3}-12a{{d}_{3}}\alpha _{1}^{2}{{\alpha}_{3}}-12a{{d}_{3}}{{\alpha}_{1}}\alpha _{3}^{2}-3a{{b}_{000}}d_{2}^{2}{{\alpha }_{3}}}{d\left( d_{2}^{2}-4{{d}_{3}} \right)}-\frac{12a{{b}_{000}}{{d}_{3}}{{\alpha}_{3}}+cd_{2}^{2}{{\alpha }_{3}}-4c{{d}_{3}}{{\alpha }_{3}}+d_{2}^{2}{{\gamma }_{3}}-4{{d}_{3}}{{\gamma }_{3}}-4a{{d}_{3}}\alpha _{3}^{3}}{d\left(d_{2}^{2}-4{{d}_{3}} \right)},
$\\
${{\gamma }_{1}}=-\frac{a\alpha _{1}^{3}d_{2}^{2}+3ad_{2}^{2}\alpha _{1}^{2}{{\alpha }_{3}}+3ad_{2}^{2}{{\alpha }_{1}}\alpha _{3}^{2}-7ad_{2}^{2}\alpha _{3}^{3}-28a{{d}_{3}}\alpha _{1}^{3}+12a{{d}_{3}}\alpha _{1}^{2}{{\alpha }_{3}}}{d_{2}^{2}-4{{d}_{3}}}-\frac{cd_{2}^{2}{{\alpha }_{1}}-cd_{2}^{2}{{\alpha }_{3}}-4c{{d}_{3}}{{\alpha }_{1}}4c{{d}_{3}}{{\alpha }_{3}}-d_{2}^{2}{{\gamma }_{3}}+4{{d}_{3}}{{\gamma }_{3}}}{d_{2}^{2}-4{{d}_{3}}}
  -\frac{12a{{d}_{3}}{{\alpha }_{1}}\alpha _{3}^{2}-3a{{b}_{000}}d_{2}^{2}{{\alpha }_{1}}4a{{d}_{3}}\alpha _{3}^{3}+3a{{b}_{000}}d_{2}^{2}{{\alpha }_{3}}12a{{b}_{000}}{{d}_{3}}{{\alpha }_{1}}-12a{{b}_{000}}{{d}_{3}}{{\alpha }_{3}}}{d_{2}^{2}-4{{d}_{3}}},\  {{\gamma }_{2}}={{\gamma }_{2}},\ {{\gamma }_{3}}={{\gamma }_{3}}.
$\\
(3) ${{a}_{000}}=\frac{{{\alpha }_{3}}\left( 4\beta _{3}^{2}-3{{c}_{000}} \right)}{3{{\beta }_{3}}}$, ${{b}_{000}}={{b}_{000}}$, ${{c}_{000}}={{c}_{000}}$, ${{\alpha }_{1}}={{\alpha }_{1}}$, ${{\alpha }_{2}}={{\alpha }_{2}}$, ${{\alpha }_{3}}=\varepsilon {{\alpha }_{1}}$, ${{\beta }_{1}}={{\beta }_{1}}$, ${{\beta }_{2}}={{\beta }_{2}}$, ${{\beta }_{3}}=\varepsilon {{\beta }_{1}}$, ${{\gamma }_{2}}={{\gamma }_{2}}$,
${{\gamma }_{1}}=-\frac{3a{{c}_{000}}\alpha _{1}^{3}-3a{{b}_{000}}{{\alpha }_{3}}\beta _{1}^{2}+c{{\alpha }_{3}}\beta _{3}^{2}+d\beta _{1}^{3}}{\beta _{3}^{2}}$, ${{\gamma }_{3}}=\varepsilon {{\gamma }_{1}}$.\\
(4) ${{a}_{000}}=0$, ${{b}_{000}}={{b}_{000}}$, ${{c}_{000}}=\frac{4\beta _{1}^{2}}{3}$, ${{\alpha }_{1}}={{\alpha }_{1}}$, ${{\alpha }_{2}}={{\alpha }_{2}}$, ${{\alpha }_{3}}={{\alpha }_{3}}$, ${{\beta }_{2}}={{\beta }_{2}}$, ${{\beta }_{3}}=-{{\beta }_{1}}$,\\
${{\beta }_{1}}=-\frac{4a{{d}_{3}}\alpha _{3}^{3}-4ad_{2}^{2}\alpha _{3}^{3}+12a{{d}_{3}}\alpha _{1}^{3}+12a{{d}_{3}}\alpha _{1}^{2}{{\alpha }_{3}}-12a{{d}_{3}}{{\alpha }_{1}}\alpha _{3}^{2}}{d\left( d_{2}^{2}-4{{d}_{3}} \right)}
  -\frac{3a{{b}_{000}}d_{2}^{2}{{\alpha }_{3}}-12a{{b}_{000}}{{d}_{3}}{{\alpha }_{3}}-cd_{2}^{2}{{\alpha }_{3}}+4c{{d}_{3}}{{\alpha }_{3}}-d_{2}^{2}{{\gamma }_{3}}+4{{d}_{3}}{{\gamma }_{3}}}{d\left( d_{2}^{2}-4{{d}_{3}} \right)}, \\
$\\
${{\gamma }_{1}}=-\frac{ad_{2}^{2}\alpha _{1}^{3}-3ad_{2}^{2}\alpha _{1}^{2}{{\alpha }_{3}}+3ad_{2}^{2}{{\alpha }_{1}}\alpha _{3}^{2}+7ad_{2}^{2}\alpha _{3}^{3}-28a{{d}_{3}}\alpha _{1}^{3}-12a{{d}_{3}}\alpha _{1}^{2}{{\alpha }_{3}}}{d_{2}^{2}-4{{d}_{3}}}-\frac{cd_{2}^{2}{{\alpha }_{1}}+cd_{2}^{2}{{\alpha }_{3}}-4c{{d}_{3}}{{\alpha }_{1}}-4c{{d}_{3}}{{\alpha }_{3}}+d_{2}^{2}{{\gamma }_{3}}-4{{d}_{3}}{{\gamma }_{3}}}{d_{2}^{2}-4{{d}_{3}}}
  -\frac{12a{{d}_{3}}{{\alpha }_{1}}\alpha _{3}^{2}-3a{{b}_{000}}d_{2}^{2}{{\alpha }_{1}}-4a{{d}_{3}}\alpha _{3}^{3}-3a{{\alpha }_{3}}{{b}_{000}}d_{2}^{2}+12a{{b}_{000}}{{d}_{3}}{{\alpha }_{1}}+12a{{b}_{000}}{{d}_{3}}{{\alpha }_{3}}}{d_{2}^{2}-4{{d}_{3}}},\ {{\gamma }_{2}}={{\gamma }_{2}},\ {{\gamma }_{3}}={{\gamma }_{3}}.
$\\
(5) ${{a}_{000}}=0$, ${{b}_{000}}={{b}_{000}}$, ${{c}_{000}}=0$, ${{\alpha }_{1}}=0$, ${{\alpha }_{2}}={{\alpha }_{2}}$, ${{\alpha }_{3}}={{\alpha }_{3}}$, ${{\beta }_{1}}={{\beta }_{1}}$, ${{\beta }_{2}}={{\beta }_{2}}$, ${{\beta }_{3}}=0$, ${{\gamma }_{1}}=-d{{\beta }_{1}}$, ${{\gamma }_{2}}={{\gamma }_{2}}$, ${{\gamma }_{3}}=3a{{b}_{000}}{{\alpha }_{3}}-a\alpha _{3}^{3}-c{{\alpha }_{3}}$.\\
Accordingly, we get ${{w}_{4}}={{e}^{-{{\xi }_{1}}}}+{{d}_{2}}\cosh {{\xi }_{3}}+{{d}_{3}}{{e}^{{{\xi }_{1}}}}$, where\\
(1) ${{\xi }_{1}}={{\alpha }_{1}}x-{{\alpha }_{1}}\left( a\alpha _{1}^{2}-3a{{b}_{000}}+c \right)t$, ${{\xi }_{3}}={{\beta }_{3}}y-d{{\beta }_{3}}t$,\\
(2) ${{\xi }_{1}}={{\alpha }_{1}}x+{{\beta }_{1}}y+{{\gamma }_{1}}t$, ${{\xi }_{3}}={{\alpha }_{3}}x+{{\beta }_{1}}y+{{\gamma }_{3}}t$,\\
(3) ${{\xi }_{1}}={{\alpha }_{1}}x+{{\beta }_{1}}y+\left( \frac{3a{{b}_{000}}{{\alpha }_{3}}\beta _{1}^{2}-3a{{c}_{000}}\alpha _{1}^{3}-c{{\alpha }_{3}}\beta _{3}^{2}-d\beta _{1}^{3}}{\beta _{3}^{2}} \right)t$,\\
${{\xi }_{3}}=\varepsilon {{\alpha }_{1}}x+{{\beta }_{1}}y+\varepsilon \left( \frac{3a{{b}_{000}}{{\alpha }_{3}}\beta _{1}^{2}-3a{{c}_{000}}\alpha _{1}^{3}-c{{\alpha }_{3}}\beta _{3}^{2}-d\beta _{1}^{3}}{\beta _{3}^{2}} \right)t$,\\
(4) ${{\xi }_{1}}={{\alpha }_{1}}x+{{\beta }_{1}}y+{{\gamma }_{1}}t$, ${{\xi }_{3}}={{\alpha }_{3}}x-{{\beta }_{1}}y+{{\gamma }_{3}}t$,\\
(5) ${{\xi }_{1}}={{\beta }_{1}}y-d{{\beta }_{1}}t$, ${{\xi }_{3}}={{\alpha }_{3}}x+\left( 3a{{b}_{000}}{{\alpha }_{3}}-a\alpha _{3}^{3}-c{{\alpha }_{3}} \right)t$.

{\bf Case 5.} ${{d}_{1}}=0$, ${{d}_{2}}={{d}_{2}}$, ${{d}_{3}}=\frac{d_{2}^{2}}{4}$.\\
(1) ${{a}_{000}}=\frac{{{\alpha }_{3}}\left( 4\beta _{3}^{2}-3{{c}_{000}} \right)}{3{{\beta }_{3}}}$, ${{b}_{000}}={{b}_{000}}$, ${{c}_{000}}={{c}_{000}}$, ${{\alpha }_{1}}={{\alpha }_{1}}$, ${{\alpha }_{2}}={{\alpha }_{2}}$, ${{\alpha }_{3}}=\varepsilon {{\alpha }_{1}}$, ${{\beta }_{1}}={{\beta }_{1}}$, ${{\beta }_{2}}={{\beta }_{2}}$, ${{\beta }_{3}}=\varepsilon {{\beta }_{1}}$, ${{\gamma }_{1}}=\frac{6a{{b}_{000}}{{\alpha }_{1}}\beta _{1}^{2}-6a{{c}_{000}}\alpha _{1}^{3}-2c{{\alpha }_{1}}\beta _{1}^{2}-2d\beta _{1}^{3}-\varepsilon \beta _{1}^{2}{{\gamma }_{3}}}{\beta _{1}^{2}}$, ${{\gamma }_{2}}={{\gamma }_{2}}$, ${{\gamma }_{3}}={{\gamma }_{3}}$.\\
(2) ${{a}_{000}}=0$, ${{b}_{000}}={{b}_{000}}$, ${{c}_{000}}={{c}_{000}}$, ${{\alpha }_{1}}={{\alpha }_{1}}$, ${{\alpha }_{2}}={{\alpha }_{2}}$, ${{\alpha }_{3}}=\varepsilon {{\alpha }_{1}}$, ${{\beta }_{1}}={{\beta }_{1}}$, ${{\beta }_{2}}={{\beta }_{2}}$, ${{\beta }_{3}}=-\varepsilon {{\beta }_{1}}$, ${{\gamma }_{1}}=\varepsilon {{\gamma }_{3}}-2d{{\beta }_{1}}$, ${{\gamma }_{2}}={{\gamma }_{2}}$, ${{\gamma }_{3}}={{\gamma }_{3}}$.\\
Accordingly, we get ${{w}_{5}}={{e}^{-{{\xi }_{1}}}}+{{d}_{2}}\cosh {{\xi }_{3}}+\frac{d_{2}^{2}}{4}{{e}^{{{\xi }_{1}}}}$, where\\
(1) ${{\xi }_{1}}={{\alpha }_{1}}x+{{\beta }_{1}}y+\left( \frac{6a{{b}_{000}}{{\alpha }_{1}}\beta _{1}^{2}-6a{{c}_{000}}\alpha _{1}^{3}-2c{{\alpha }_{1}}\beta _{1}^{2}-2d\beta _{1}^{3}-\varepsilon \beta _{1}^{2}{{\gamma }_{3}}}{\beta _{1}^{2}} \right)t$,
${{\xi }_{3}}=\varepsilon {{\alpha }_{3}}x+\varepsilon {{\beta }_{3}}y+{{\gamma }_{3}}t$,\\
(2) ${{\xi }_{1}}={{\alpha }_{1}}x+{{\beta }_{1}}y+\left( \varepsilon {{\gamma }_{3}}-2d{{\beta }_{1}} \right)t$, ${{\xi }_{3}}=\varepsilon {{\alpha }_{1}}x-\varepsilon {{\beta }_{1}}y+{{\gamma }_{3}}t$.

{\bf Case 6.} ${{d}_{1}}=0$, ${{d}_{2}}={{d}_{2}}$.\\
(1) ${{d}_{3}}=\frac{{{\beta }_{3}}d_{2}^{2}\left( 2{{\alpha }_{1}}{{\beta }_{1}}{{\beta }_{3}}+{{\alpha }_{3}}\beta _{1}^{2}-3{{\alpha }_{3}}\beta _{3}^{2} \right)}{4{{\beta }_{1}}\left( 3{{\alpha }_{1}}\beta _{1}^{2}-{{\alpha }_{1}}\beta _{3}^{2}-2{{\alpha }_{3}}{{\beta }_{1}}{{\beta }_{3}} \right)}$, ${{a}_{000}}=\frac{2{{\beta }_{1}}{{\beta }_{3}}\left( \alpha _{1}^{2}-\alpha _{3}^{2} \right)}{3\left( {{\alpha }_{1}}{{\beta }_{3}}-{{\alpha }_{3}}{{\beta }_{1}} \right)}$, ${{c}_{000}}=\frac{{{\alpha }_{1}}\beta _{1}^{2}{{\beta }_{3}}-{{\alpha }_{1}}\beta _{3}^{3}+{{\alpha }_{3}}\beta _{1}^{3}-{{\alpha }_{3}}{{\beta }_{1}}\beta _{3}^{2}}{-3\left( {{\alpha }_{1}}{{\beta }_{3}}-{{\alpha }_{3}}{{\beta }_{1}} \right)}$,\\
${{b}_{000}}=\frac{2a\alpha _{1}^{4}{{\beta }_{1}}\beta _{3}^{2}-a\alpha _{1}^{3}{{\alpha }_{3}}\beta _{1}^{2}{{\beta }_{3}}-3a\alpha _{1}^{3}{{\alpha }_{3}}\beta _{3}^{3}-3a\alpha _{1}^{2}\alpha _{3}^{2}\beta _{1}^{3}+3a\alpha _{1}^{2}\alpha _{3}^{2}{{\beta }_{1}}\beta _{3}^{2}+5a{{\alpha }_{1}}\alpha _{3}^{3}\beta _{1}^{2}{{\beta }_{3}}}{3a{{\alpha }_{3}}\left( {{\alpha }_{1}}{{\beta }_{3}}-{{\alpha }_{3}}{{\beta }_{1}} \right)\left( \beta _{1}^{2}-\beta _{3}^{2} \right)}+ \\
\frac{c{{\alpha }_{1}}{{\alpha }_{3}}\beta _{1}^{2}{{\beta }_{3}}-c{{\alpha }_{1}}{{\alpha }_{3}}\beta _{3}^{3}+d{{\alpha }_{1}}\beta _{1}^{2}\beta _{3}^{2}-d{{\alpha }_{1}}\beta _{3}^{4}-c\alpha _{3}^{2}\beta _{1}^{3}-a\alpha _{3}^{4}\beta _{1}^{3}-a\alpha _{3}^{4}{{\beta }_{1}}\beta _{3}^{2}-a{{\alpha }_{1}}\alpha _{3}^{3}\beta _{3}^{3}}{3a{{\alpha }_{3}}\left( {{\alpha }_{1}}{{\beta }_{3}}-{{\alpha }_{3}}{{\beta }_{1}} \right)\left( \beta _{1}^{2}-\beta _{3}^{2} \right)}+\\
\frac{c\alpha _{3}^{2}{{\beta }_{1}}\beta _{3}^{2}-d{{\alpha }_{3}}\beta _{1}^{3}{{\beta }_{3}}+d{{\alpha }_{3}}{{\beta }_{1}}\beta _{3}^{3}+{{\alpha }_{1}}\beta _{1}^{2}{{\beta }_{3}}{{\gamma }_{3}}-{{\alpha }_{1}}\beta _{3}^{3}{{\gamma }_{3}}-{{\alpha }_{3}}\beta _{1}^{3}{{\gamma }_{3}}+{{\alpha }_{3}}{{\beta }_{1}}\beta _{3}^{2}{{\gamma }_{3}}}{3a{{\alpha }_{3}}\left( {{\alpha }_{1}}{{\beta }_{3}}-{{\alpha }_{3}}{{\beta }_{1}} \right)\left( \beta _{1}^{2}-\beta _{3}^{2} \right)},$ $
{{\alpha }_{1}}={{\alpha }_{1}}$, ${{\alpha }_{2}}={{\alpha }_{2}}$, ${{\alpha }_{3}}={{\alpha }_{3}}$, ${{\beta }_{1}}={{\beta }_{1}}$, ${{\beta }_{2}}={{\beta }_{2}}$, ${{\beta }_{3}}={{\beta }_{3}}$, ${{\gamma }_{2}}={{\gamma }_{2}}$, ${{\gamma }_{3}}={{\gamma }_{3}}$,
${{\gamma }_{1}}=\frac{2a\alpha _{1}^{4}{{\beta }_{1}}{{\beta }_{3}}+2a\alpha _{1}^{3}{{\alpha }_{3}}\beta _{1}^{2}-2a\alpha _{1}^{3}{{\alpha }_{3}}\beta _{3}^{2}-4a\alpha _{1}^{2}\alpha _{3}^{2}{{\beta }_{1}}{{\beta }_{3}}-2a{{\alpha }_{1}}\alpha _{3}^{3}\beta _{1}^{2}+2a{{\alpha }_{1}}\alpha _{3}^{3}\beta _{3}^{2}}{{{\alpha }_{3}}\left( \beta _{1}^{2}-\beta _{3}^{2} \right)}+\\
\frac{2a\alpha _{3}^{4}{{\beta }_{1}}{{\beta }_{3}}+d{{\alpha }_{1}}\beta _{1}^{2}{{\beta }_{3}}-d{{\alpha }_{1}}\beta _{3}^{3}-d{{\alpha }_{3}}\beta _{1}^{3}+d{{\alpha }_{3}}{{\beta }_{1}}\beta _{3}^{2}+{{\alpha }_{1}}\beta _{1}^{2}{{\gamma }_{3}}-{{\alpha }_{1}}\beta _{3}^{2}{{\gamma }_{3}}}{{{\alpha }_{3}}\left( \beta _{1}^{2}-\beta _{3}^{2} \right)}. \\
$\\
(2) ${{d}_{3}}=\frac{d_{2}^{2}\beta _{3}^{2}}{2\left( 3\beta _{1}^{2}-\beta _{3}^{2} \right)}$, ${{a}_{000}}=\frac{2{{\alpha }_{1}}{{\beta }_{1}}}{3}$, ${{b}_{000}}={{b}_{000}}$, ${{c}_{000}}=\frac{\beta _{3}^{2}-\beta _{1}^{2}}{3}$, ${{\alpha }_{1}}={{\alpha }_{1}}$, ${{\alpha }_{2}}={{\alpha }_{2}}$, ${{\alpha }_{3}}=0$, ${{\beta }_{1}}={{\beta }_{1}}$, ${{\beta }_{2}}={{\beta }_{2}}$, ${{\beta }_{3}}={{\beta }_{3}}$, ${{\gamma }_{1}}=\frac{a\alpha _{1}^{3}\beta _{1}^{2}+a\alpha _{1}^{3}\beta _{3}^{2}+3a{{b}_{000}}{{\alpha }_{1}}\beta _{1}^{2}-3a{{b}_{000}}{{\alpha }_{1}}\beta _{3}^{2}-c{{\alpha }_{1}}\beta _{1}^{2}+c{{\alpha }_{1}}\beta _{3}^{2}-d\beta _{1}^{3}+d{{\beta }_{1}}\beta _{3}^{2}}{\left( \beta _{1}^{2}-\beta _{3}^{2} \right)}$, ${{\gamma }_{2}}={{\gamma }_{2}}$, ${{\gamma }_{3}}=\frac{{{\beta }_{3}}\left( 2a\alpha _{1}^{3}{{\beta }_{1}}+d\beta _{1}^{2}-d\beta _{3}^{2} \right)}{-\left( \beta _{1}^{2}-\beta _{3}^{2} \right)}$.
\\
Accordingly, we get ${{w}_{6}}={{e}^{-{{\xi }_{1}}}}+{{d}_{2}}\cosh {{\xi }_{3}}+{{d}_{3}}{{e}^{{{\xi }_{1}}}}$, where\\
(1) ${{\xi }_{1}}={{\alpha }_{1}}x+{{\beta }_{1}}y+{{\gamma }_{1}}t$, ${{\xi }_{3}}={{\alpha }_{3}}x+{{\beta }_{3}}y+{{\gamma }_{3}}t$,\\ (2) ${{\xi }_{1}}={{\alpha }_{1}}x+{{\beta }_{1}}y+{{\gamma }_{1}}t$, ${{\xi }_{3}}={{\beta }_{3}}y+{{\gamma }_{3}}t$.

{\bf Case 7.} ${{d}_{1}}={{d}_{1}}$, ${{d}_{2}}=0$, ${{d}_{3}}={{d}_{3}}$.\\
(1) ${{a}_{000}}=0$, ${{b}_{000}}={{b}_{000}}$, ${{c}_{000}}=-\frac{\beta _{2}^{2}}{3}$, ${{\alpha }_{1}}={{\alpha }_{1}}$, ${{\alpha }_{2}}=0$, ${{\alpha }_{3}}={{\alpha }_{3}}$, ${{\beta }_{1}}=0$, ${{\beta }_{2}}={{\beta }_{2}}$, ${{\beta }_{3}}={{\beta }_{3}}$, ${{\gamma }_{1}}=-{{\alpha }_{1}}\left( a\alpha _{1}^{2}-3a{{b}_{000}}+c \right)$, ${{\gamma }_{2}}=-d{{\beta }_{2}}$, ${{\gamma }_{3}}={{\gamma }_{3}}$.\\
(2) ${{a}_{000}}=0$, ${{b}_{000}}={{b}_{000}}$, ${{c}_{000}}=\frac{\beta _{1}^{2}}{3}$, ${{\alpha }_{1}}=0$, ${{\alpha }_{2}}={{\alpha }_{2}}$, ${{\alpha }_{3}}={{\alpha }_{3}}$, ${{\beta }_{1}}={{\beta }_{1}}$, ${{\beta }_{2}}=0$, ${{\beta }_{3}}={{\beta }_{3}}$, ${{\gamma }_{1}}=-d{{\beta }_{1}}$, ${{\gamma }_{2}}=a\alpha _{2}^{3}+3a{{\alpha }_{2}}{{b}_{000}}-c{{\alpha }_{2}}$, ${{\gamma }_{3}}={{\gamma }_{3}}$.\\
(3) ${{a}_{000}}=0$, ${{c}_{000}}=-\frac{4\beta _{2}^{2}}{3}$, ${{\alpha }_{1}}={{\alpha }_{1}}$, ${{\alpha }_{2}}={{\alpha }_{2}}$, ${{\alpha }_{3}}={{\alpha }_{3}}$, ${{\beta }_{1}}=\varepsilon I{{\beta }_{2}}$, ${{\beta }_{2}}={{\beta }_{2}}$, ${{\beta }_{3}}={{\beta }_{3}}$, ${{\gamma }_{2}}={{\gamma }_{2}}$, ${{\gamma }_{3}}={{\gamma }_{3}}$,\\
${{b}_{000}}=\frac{4a\alpha _{2}^{3}d_{1}^{2}+12a{{d}_{3}}\alpha _{1}^{2}{{\alpha }_{2}}-4a{{d}_{3}}\alpha _{2}^{3}-cd_{1}^{2}{{\alpha }_{2}}-dd_{1}^{2}{{\beta }_{2}}+4c{{d}_{3}}{{\alpha }_{2}}+4d{{d}_{3}}{{\beta }_{2}}-d_{1}^{2}{{\gamma }_{2}}+4{{d}_{3}}{{\gamma }_{2}}}{3a\left( 4{{d}_{3}}-d_{1}^{2} \right){{\alpha }_{2}}}
  +\frac{\varepsilon I\left( 4a{{d}_{3}}\alpha _{1}^{3}+4a{{d}_{3}}{{\alpha }_{1}}\alpha _{2}^{2} \right)}{a\left( 4{{d}_{3}}-d_{1}^{2} \right){{\alpha }_{2}}}, \
 {{\gamma }_{1}}=-\frac{{{\alpha }_{1}}\left( a\alpha _{1}^{2}{{\alpha }_{2}}+a\alpha _{2}^{3}-d{{\beta }_{2}}-{{\gamma }_{2}} \right)}{{{\alpha }_{2}}}
  +\frac{\varepsilon I\left( 3ad_{1}^{2}\alpha _{1}^{2}\alpha _{2}^{2}+3ad_{1}^{2}\alpha _{2}^{4}+12a{{d}_{3}}\alpha _{1}^{4}+12a{{d}_{3}}\alpha _{1}^{2}\alpha _{2}^{2}+dd_{1}^{2}{{\alpha }_{2}}{{\beta }_{2}}-4d{{d}_{3}}{{\alpha }_{2}}{{\beta }_{2}} \right)}{\left( 4{{d}_{3}}-d_{1}^{2} \right){{\alpha }_{2}}}.
$\\
(4) ${{A}^{6}}=\frac{{{\left( d_{1}^{2}-4{{d}_{3}} \right)}^{2}}{{\left( d{{\beta }_{2}}+{{\gamma }_{2}} \right)}^{2}}}{-24{{a}^{2}}{{d}_{3}}\left( d_{1}^{2}+2{{d}_{3}} \right)}$,\  ${{a}_{000}}=\frac{\left( d_{1}^{2}-4{{d}_{3}} \right)\left( d{{\beta }_{2}}+{{\gamma }_{2}} \right){{\beta }_{2}}}{18a{{A}^{2}}{{d}_{3}}}$, ${{b}_{000}}={{b}_{000}}$, ${{c}_{000}}=\frac{\beta _{2}^{2}\left( d_{1}^{2}-4{{d}_{3}} \right)}{18{{d}_{3}}}$, ${{\alpha }_{1}}=A$, ${{\alpha }_{2}}=0$, ${{\alpha }_{3}}={{\alpha }_{3}}$, ${{\beta }_{1}}=\frac{\left( d_{1}^{2}-4{{d}_{3}} \right)\left( d{{\beta }_{2}}+{{\gamma }_{2}} \right){{\beta }_{2}}}{12a{{A}^{3}}{{d}_{3}}}$, ${{\beta }_{2}}={{\beta }_{2}}$, ${{\beta }_{3}}={{\beta }_{3}}$, ${{\gamma }_{2}}={{\gamma }_{2}}$, ${{\gamma }_{3}}={{\gamma }_{3}}$,\\
${{\gamma }_{1}}=\frac{\left( 72{{a}^{2}}{{b}_{000}}d_{1}^{2}{{d}_{3}}+144{{a}^{2}}{{b}_{000}}d_{3}^{2}-24acd_{1}^{2}{{d}_{3}}-48acd_{3}^{2} \right){{A}^{4}}-3{{d}^{2}}d_{1}^{4}\beta _{2}^{2}-4dd_{1}^{4}{{\beta }_{2}}{{\gamma }_{2}}}{24a{{d}_{3}}{{A}^{3}}\left( d_{1}^{2}+2{{d}_{3}} \right)} \\
  +\frac{48{{d}^{2}}d_{3}^{2}\beta _{2}^{2}-4dd_{1}^{2}{{d}_{3}}{{\beta }_{2}}{{\gamma }_{2}}-d_{1}^{4}\gamma _{2}^{2}+80dd_{3}^{2}{{\beta }_{2}}{{\gamma }_{2}}-4d_{1}^{2}{{d}_{3}}\gamma _{2}^{2}+32d_{3}^{2}\gamma _{2}^{2}}{24a{{d}_{3}}{{A}^{3}}\left( d_{1}^{2}+2{{d}_{3}} \right)}.
$\\
(5) ${{a}_{000}}=0$, ${{b}_{000}}=\frac{4a\alpha _{2}^{3}-c{{\alpha }_{2}}-d{{\beta }_{2}}-{{\gamma }_{2}}}{-3a{{\alpha }_{2}}}$, ${{c}_{000}}=-\frac{4\beta _{2}^{2}}{3}$, ${{\alpha }_{1}}=-\varepsilon I{{\alpha }_{2}}$, ${{\alpha }_{2}}={{\alpha }_{2}}$, ${{\alpha }_{3}}={{\alpha }_{3}}$, ${{\beta }_{1}}=\varepsilon I{{\beta }_{2}}$, ${{\beta }_{2}}={{\beta }_{2}}$, ${{\beta }_{3}}={{\beta }_{3}}$, ${{\gamma }_{1}}=-\varepsilon I\left( 2d{{\beta }_{2}}+{{\gamma }_{2}} \right)$, ${{\gamma }_{2}}={{\gamma }_{2}}$, ${{\gamma }_{3}}={{\gamma }_{3}}$.\\
(6) ${{a}_{000}}=\frac{{{\alpha }_{2}}\left( 4\beta _{2}^{2}+3{{c}_{000}} \right)}{-3{{\beta }_{2}}}$, ${{b}_{000}}=\frac{3a{{c}_{000}}\alpha _{2}^{3}+c{{\alpha }_{2}}\beta _{2}^{2}+d\beta _{2}^{3}+\beta _{2}^{2}{{\gamma }_{2}}}{3a{{\alpha }_{2}}\beta _{2}^{2}}$, ${{c}_{000}}={{c}_{000}}$, ${{\alpha }_{1}}=\varepsilon I{{\alpha }_{2}}$, ${{\alpha }_{2}}={{\alpha }_{2}}$, ${{\alpha }_{3}}={{\alpha }_{3}}$, ${{\beta }_{1}}=\varepsilon I{{\beta }_{2}}$, ${{\beta }_{2}}={{\beta }_{2}}$, ${{\beta }_{3}}={{\beta }_{3}}$, ${{\gamma }_{1}}=\varepsilon I{{\gamma }_{2}}$, ${{\gamma }_{2}}={{\gamma }_{2}}$, ${{\gamma }_{3}}={{\gamma }_{3}}$.\\
(7) ${{B}^{6}}=\frac{{{\left( d_{1}^{2}-4{{d}_{3}} \right)}^{2}}{{\left( d{{\beta }_{2}}+{{\gamma }_{2}} \right)}^{2}}}{-144{{a}^{2}}d_{3}^{2}}$, ${{a}_{000}}=0$, ${{b}_{000}}={{b}_{000}}$, ${{c}_{000}}=-\frac{4\beta _{2}^{2}}{3}$, ${{\alpha }_{1}}=B$, ${{\alpha }_{2}}=0$, ${{\alpha }_{3}}={{\alpha }_{3}}$, ${{\beta }_{1}}=\frac{{{\beta }_{2}}\left( d_{1}^{2}-4{{d}_{3}} \right)\left( d{{\beta }_{2}}+{{\gamma }_{2}} \right)}{12a{{d}_{3}}{{B}^{3}}}$, ${{\beta }_{2}}={{\beta }_{2}}$, ${{\beta }_{3}}={{\beta }_{3}}$, ${{\gamma }_{2}}={{\gamma }_{2}}$, ${{\gamma }_{3}}={{\gamma }_{3}}$,
${{\gamma }_{1}}=\frac{\left( 432{{a}^{2}}{{b}_{000}}d_{3}^{2}-144acd_{3}^{2} \right){{B}^{4}}+{{d}^{2}}d_{1}^{4}\beta _{2}^{2}-32{{d}^{2}}d_{1}^{2}{{d}_{3}}\beta _{2}^{2}+2dd_{1}^{4}{{\beta }_{2}}{{\gamma }_{2}}}{144ad_{3}^{2}{{B}^{3}}}
  +\frac{112{{d}^{2}}d_{3}^{2}\beta _{2}^{2}-52dd_{1}^{2}{{d}_{3}}{{\beta }_{2}}{{\gamma }_{2}}+d_{1}^{4}\gamma _{2}^{2}+176dd_{3}^{2}{{\beta }_{2}}{{\gamma }_{2}}-20d_{1}^{2}{{d}_{3}}\gamma _{2}^{2}+64d_{3}^{2}\gamma _{2}^{2}}{144ad_{3}^{2}{{B}^{3}}}.
$\\
Accordingly, we get ${{w}_{7}}={{e}^{-{{\xi }_{1}}}}+{{d}_{1}}\cos {{\xi }_{2}}+{{d}_{3}}{{e}^{{{\xi }_{1}}}}$, where\\
(1) ${{\xi }_{1}}={{\alpha }_{1}}x-{{\alpha }_{1}}\left( a\alpha _{1}^{2}-3a{{b}_{000}}+c \right)t$, ${{\xi }_{2}}={{\beta }_{2}}y-d{{\beta }_{2}}t$,\\
(2) ${{\xi }_{1}}={{\beta }_{1}}y-d{{\beta }_{1}}t$, ${{\xi }_{2}}={{\alpha }_{2}}x+\left( a\alpha _{2}^{3}+3a{{\alpha }_{2}}{{b}_{000}}-c{{\alpha }_{2}} \right)t$,\\
(3) ${{\xi }_{1}}={{\alpha }_{1}}x+\varepsilon I{{\beta }_{2}}y+{{\gamma }_{1}}t$, ${{\xi }_{2}}={{\alpha }_{2}}x+{{\beta }_{2}}y+{{\gamma }_{2}}t$,\\
(4) ${{\xi }_{1}}=Ax+{{\beta }_{1}}y+{{\gamma }_{1}}t$, ${{\xi }_{2}}={{\beta }_{2}}y+{{\gamma }_{2}}t$,\\
(5) ${{\xi }_{1}}=-\varepsilon I{{\alpha }_{2}}x+\varepsilon I{{\beta }_{2}}y-\varepsilon I\left( 2d{{\beta }_{2}}+{{\gamma }_{2}} \right)t$, ${{\xi }_{2}}={{\alpha }_{2}}x+{{\beta }_{2}}y+{{\gamma }_{2}}t$,\\
(6) ${{\xi }_{1}}=\varepsilon I{{\alpha }_{2}}x+\varepsilon I{{\beta }_{2}}y+\varepsilon I{{\gamma }_{2}}t$, ${{\xi }_{2}}={{\alpha }_{2}}x+{{\beta }_{2}}y+{{\gamma }_{2}}t$,\\
(7) ${{\xi }_{1}}=Bx+{{\beta }_{1}}y+{{\gamma }_{1}}t$, ${{\xi }_{2}}={{\beta }_{2}}y+{{\gamma }_{2}}t$.

{\bf Case 8.} ${{d}_{1}}={{d}_{1}}$, ${{d}_{2}}=0$.\\
(1) ${{d}_{3}}=\frac{d_{1}^{2}{{\beta }_{2}}\left( 2{{\alpha }_{1}}{{\beta }_{1}}{{\beta }_{2}}+{{\alpha }_{2}}\beta _{1}^{2}+3{{\alpha }_{2}}\beta _{2}^{2} \right)}{-4{{\beta }_{1}}\left( 3{{\alpha }_{1}}\beta _{1}^{2}+{{\alpha }_{1}}\beta _{2}^{2}+2{{\alpha }_{2}}{{\beta }_{1}}{{\beta }_{2}} \right)}$, ${{a}_{000}}=\frac{2{{\beta }_{1}}{{\beta }_{2}}\left( \alpha _{1}^{2}+\alpha _{2}^{2} \right)}{3\left( {{\alpha }_{1}}{{\beta }_{2}}-{{\alpha }_{2}}{{\beta }_{1}} \right)}$, ${{c}_{000}}=\frac{{{\alpha }_{1}}\beta _{1}^{2}{{\beta }_{2}}+{{\alpha }_{1}}\beta _{2}^{3}+{{\alpha }_{2}}\beta _{1}^{3}+{{\alpha }_{2}}{{\beta }_{1}}\beta _{2}^{2}}{-3\left( {{\alpha }_{1}}{{\beta }_{2}}-{{\alpha }_{2}}{{\beta }_{1}} \right)}$,\\
${{b}_{000}}=\frac{2a\alpha _{1}^{4}{{\beta }_{1}}\beta _{2}^{2}-a\alpha _{1}^{3}{{\alpha }_{2}}\beta _{1}^{2}{{\beta }_{2}}+3a\alpha _{1}^{3}{{\alpha }_{2}}\beta _{2}^{3}-3a\alpha _{1}^{2}\alpha _{2}^{2}\beta _{1}^{3}-3a\alpha _{1}^{2}\alpha _{2}^{2}{{\beta }_{1}}\beta _{2}^{2}+c{{\alpha }_{1}}{{\alpha }_{2}}\beta _{1}^{2}{{\beta }_{2}}}{3a{{\alpha }_{2}}\left( {{\alpha }_{1}}{{\beta }_{2}}-{{\alpha }_{2}}{{\beta }_{1}} \right)\left( \beta _{1}^{2}+\beta _{2}^{2} \right)}+\\
\frac{c{{\alpha }_{1}}{{\alpha }_{2}}\beta _{2}^{3}+d{{\alpha }_{1}}\beta _{1}^{2}\beta _{2}^{2}-a{{\alpha }_{1}}\alpha _{2}^{3}\beta _{2}^{3}+a\alpha _{2}^{4}\beta _{1}^{3}-a\alpha _{2}^{4}{{\beta }_{1}}\beta _{2}^{2}-{{\alpha }_{2}}{{\beta }_{1}}\beta _{2}^{2}{{\gamma }_{2}}-5a{{\alpha }_{1}}\alpha _{2}^{3}\beta _{1}^{2}{{\beta }_{2}}}{3a{{\alpha }_{2}}\left( {{\alpha }_{1}}{{\beta }_{2}}-{{\alpha }_{2}}{{\beta }_{1}} \right)\left( \beta _{1}^{2}+\beta _{2}^{2} \right)}+\\
\frac{d{{\alpha }_{1}}\beta _{2}^{4}-c\alpha _{2}^{2}\beta _{1}^{3}-c\alpha _{2}^{2}{{\beta }_{1}}\beta _{2}^{2}-d{{\alpha }_{2}}\beta _{1}^{3}{{\beta }_{2}}-d{{\alpha }_{2}}{{\beta }_{1}}\beta _{2}^{3}+{{\alpha }_{1}}\beta _{1}^{2}{{\beta }_{2}}{{\gamma }_{2}}+{{\alpha }_{1}}\beta _{2}^{3}{{\gamma }_{2}}-{{\alpha }_{2}}\beta _{1}^{3}{{\gamma }_{2}}}{3a{{\alpha }_{2}}\left( {{\alpha }_{1}}{{\beta }_{2}}-{{\alpha }_{2}}{{\beta }_{1}} \right)\left( \beta _{1}^{2}+\beta _{2}^{2} \right)},$ $
{{\alpha }_{1}}={{\alpha }_{1}}$, ${{\alpha }_{2}}={{\alpha }_{2}}$, ${{\alpha }_{3}}={{\alpha }_{3}}$, ${{\beta }_{1}}={{\beta }_{1}}$, ${{\beta }_{2}}={{\beta }_{2}}$, ${{\beta }_{3}}={{\beta }_{3}}$, ${{\gamma }_{2}}={{\gamma }_{2}}$, ${{\gamma }_{3}}={{\gamma }_{3}}$,
${{\gamma }_{1}}=\frac{2a\alpha _{1}^{4}{{\beta }_{1}}{{\beta }_{2}}+2a\alpha _{1}^{3}{{\alpha }_{2}}\beta _{1}^{2}+2a\alpha _{1}^{3}{{\alpha }_{2}}\beta _{2}^{2}+4a\alpha _{1}^{2}\alpha _{2}^{2}{{\beta }_{1}}{{\beta }_{2}}+2a{{\alpha }_{1}}\alpha _{2}^{3}\beta _{1}^{2}+2a{{\alpha }_{1}}\alpha _{2}^{3}\beta _{2}^{2}}{{{\alpha }_{2}}\left( \beta _{1}^{2}+\beta _{2}^{2} \right)}+\\
  \frac{2a\alpha _{2}^{4}{{\beta }_{1}}{{\beta }_{2}}+d{{\alpha }_{1}}\beta _{1}^{2}{{\beta }_{2}}+d{{\alpha }_{1}}\beta _{2}^{3}-d{{\alpha }_{2}}\beta _{1}^{3}-d{{\alpha }_{2}}{{\beta }_{1}}\beta _{2}^{2}+{{\alpha }_{1}}\beta _{1}^{2}{{\gamma }_{2}}+{{\alpha }_{1}}\beta _{2}^{2}{{\gamma }_{2}}}{{{\alpha }_{2}}\left( \beta _{1}^{2}+\beta _{2}^{2} \right)}. \\
$\\
(2) ${{d}_{3}}=\frac{d_{1}^{2}}{4}$, ${{a}_{000}}=0$, ${{b}_{000}}={{b}_{000}}$, ${{c}_{000}}={{c}_{000}}$, ${{\alpha }_{1}}=-\varepsilon I{{\alpha }_{2}}$, ${{\alpha }_{2}}={{\alpha }_{2}}$, ${{\alpha }_{3}}={{\alpha }_{3}}$, ${{\beta }_{1}}=\varepsilon I{{\beta }_{2}}$, ${{\beta }_{2}}={{\beta }_{2}}$, ${{\beta }_{3}}={{\beta }_{3}}$, ${{\gamma }_{1}}=-\varepsilon I\left( 2d{{\beta }_{2}}+{{\gamma }_{2}} \right)$, ${{\gamma }_{2}}={{\gamma }_{2}}$, ${{\gamma }_{3}}={{\gamma }_{3}}$.\\
(3) ${{d}_{3}}=\frac{d_{1}^{2}}{4}$, ${{a}_{000}}=\frac{{{\alpha }_{2}}\left( 4\beta _{2}^{2}+3{{c}_{000}} \right)}{-3{{\beta }_{2}}}$, ${{b}_{000}}={{b}_{000}}$, ${{c}_{000}}={{c}_{000}}$, ${{\alpha }_{1}}=\varepsilon I{{\alpha }_{2}}$, ${{\alpha }_{2}}={{\alpha }_{2}}$, ${{\alpha }_{3}}={{\alpha }_{3}}$, ${{\beta }_{1}}=\varepsilon I{{\beta }_{2}}$, ${{\beta }_{2}}={{\beta }_{2}}$, ${{\beta }_{3}}={{\beta }_{3}}$, ${{\gamma }_{1}}=\frac{\varepsilon I\left( 6a{{c}_{000}}\alpha _{2}^{3}-6a{{b}_{000}}{{\alpha }_{2}}\beta _{2}^{2}+2c{{\alpha }_{2}}\beta _{2}^{2}+2d\beta _{2}^{3}+\beta _{2}^{2}{{\gamma }_{2}} \right)}{-\beta _{2}^{2}}$, ${{\gamma }_{2}}={{\gamma }_{2}}$, ${{\gamma }_{3}}={{\gamma }_{3}}$.\\
Accordingly, we get ${{w}_{8}}={{e}^{-{{\xi }_{1}}}}+{{d}_{1}}\cos {{\xi }_{2}}+{{d}_{3}}{{e}^{{{\xi }_{1}}}}$, where\\
(1) ${{\xi }_{1}}={{\alpha }_{1}}x+{{\beta }_{1}}y+{{\gamma }_{1}}t$, ${{\xi }_{2}}={{\alpha }_{2}}x+{{\beta }_{2}}y+{{\gamma }_{2}}t$,\\
(2) ${{\xi }_{1}}=-\varepsilon I{{\alpha }_{2}}x+\varepsilon I{{\beta }_{2}}y-\varepsilon I\left( 2d{{\beta }_{2}}+{{\gamma }_{2}} \right)t$, ${{\xi }_{2}}={{\alpha }_{2}}x+{{\beta }_{2}}y+{{\gamma }_{2}}t$,\\
(3) ${{\xi }_{1}}=\varepsilon I{{\alpha }_{2}}x+\varepsilon I{{\beta }_{2}}y-\frac{\varepsilon I\left( 6a{{c}_{000}}\alpha _{2}^{3}-6a{{b}_{000}}{{\alpha }_{2}}\beta _{2}^{2}+2c{{\alpha }_{2}}\beta _{2}^{2}+2d\beta _{2}^{3}+\beta _{2}^{2}{{\gamma }_{2}} \right)t}{\beta _{2}^{2}}$,
${{\xi }_{2}}={{\alpha }_{2}}x+{{\beta }_{2}}y+{{\gamma }_{2}}t$.

{\bf Case 9.} ${{d}_{1}}={{d}_{1}}$, ${{d}_{2}}={{d}_{2}}$, ${{d}_{3}}=0$.\\
(1) ${{a}_{000}}=0$, ${{b}_{000}}={{b}_{000}}$, ${{c}_{000}}=\frac{\beta _{2}^{2}}{3}$, ${{\alpha }_{1}}={{\alpha }_{1}}$, ${{\alpha }_{2}}=\varepsilon I{{\alpha }_{3}}$, ${{\alpha }_{3}}={{\alpha }_{3}}$, ${{\beta }_{1}}={{\beta }_{1}}$, ${{\beta }_{2}}={{\beta }_{3}}=0$, ${{\gamma }_{1}}=3a{{b}_{000}}{{\alpha }_{1}}-a\alpha _{1}^{3}-3a{{\alpha }_{1}}\alpha _{3}^{2}-c{{\alpha }_{1}}-d{{\beta }_{1}}$,
${{\gamma }_{2}}=\varepsilon I{{\alpha }_{3}}\left( 3a{{b}_{000}}-3a\alpha _{1}^{2}-a\alpha _{3}^{2}-c \right)$, ${{\gamma }_{3}}=3a{{b}_{000}}{{\alpha }_{3}}-3a\alpha _{1}^{2}{{\alpha }_{3}}-a\alpha _{3}^{3}-c{{\alpha }_{3}}$.\\
(2) ${{a}_{000}}=\frac{{{\alpha }_{3}}\left( 4\beta _{3}^{2}-3{{c}_{000}} \right)}{3{{\beta }_{3}}}$, ${{b}_{000}}={{b}_{000}}$, ${{c}_{000}}={{c}_{000}}$, ${{\alpha }_{1}}={{\alpha }_{1}}$, ${{\alpha }_{2}}=\varepsilon I{{\alpha }_{1}}$, ${{\alpha }_{3}}=\varepsilon {{\alpha }_{1}}$, ${{\beta }_{1}}={{\beta }_{1}}$, ${{\beta }_{2}}=\varepsilon I{{\beta }_{1}}$, ${{\beta }_{3}}=\varepsilon {{\beta }_{1}}$, ${{\gamma }_{1}}={{\gamma }_{3}}=\frac{3a{{c}_{000}}\alpha _{1}^{3}-3a{{b}_{000}}{{\alpha }_{1}}\beta _{1}^{2}+c{{\alpha }_{1}}\beta _{1}^{2}+d\beta _{1}^{3}}{-\beta _{1}^{2}}$,
${{\gamma }_{2}}=\varepsilon I{{\gamma }_{1}}$.\\
(3) ${{a}_{000}}=0$, ${{b}_{000}}={{b}_{000}}$, ${{c}_{000}}=\frac{4\beta _{3}^{2}}{3}$, ${{\alpha }_{1}}={{\alpha }_{1}}$, ${{\alpha }_{2}}={{\alpha }_{3}}=0$, ${{\beta }_{1}}={{\beta }_{1}}$, ${{\beta }_{2}}=-\varepsilon I{{\beta }_{1}}$, ${{\beta }_{3}}=-{{\beta }_{1}}$, ${{\gamma }_{1}}=3a{{b}_{000}}{{\alpha }_{1}}-a\alpha _{1}^{3}-c{{\alpha }_{1}}-d{{\beta }_{1}}$, ${{\gamma }_{2}}=\varepsilon Id{{\beta }_{1}}$, ${{\gamma }_{3}}=d{{\beta }_{1}}$.\\
Accordingly, we get ${{w}_{9}}={{e}^{-{{\xi }_{1}}}}+{{d}_{1}}\cos {{\xi }_{2}}+{{d}_{2}}\cosh {{\xi }_{3}}$, where\\
(1) ${{\xi }_{1}}={{\alpha }_{1}}x+{{\beta }_{1}}y+\left( 3a{{b}_{000}}{{\alpha }_{1}}-a\alpha _{1}^{3}-3a{{\alpha }_{1}}\alpha _{3}^{2}-c{{\alpha }_{1}}-d{{\beta }_{1}} \right)t$,\\
${{\xi }_{2}}=\varepsilon I{{\alpha }_{3}}x+\varepsilon I{{\alpha }_{3}}\left( 3a{{b}_{000}}-3a\alpha _{1}^{2}-a\alpha _{3}^{2}-c \right)t$,
${{\xi }_{3}}={{\alpha }_{3}}x+\left( 3a{{b}_{000}}{{\alpha }_{3}}-3a\alpha _{1}^{2}{{\alpha }_{3}}-a\alpha _{3}^{3}-c{{\alpha }_{3}} \right)t$,\\
(2) ${{\xi }_{1}}={{\alpha }_{1}}x+{{\beta }_{1}}y-\left( \frac{3a{{c}_{000}}\alpha _{1}^{3}-3a{{b}_{000}}{{\alpha }_{1}}\beta _{1}^{2}+c{{\alpha }_{1}}\beta _{1}^{2}+d\beta _{1}^{3}}{\beta _{1}^{2}} \right)t$,\\
${{\xi }_{2}}=\varepsilon I{{\alpha }_{1}}x+\varepsilon I{{\beta }_{1}}y-\varepsilon I\left( \frac{3a{{c}_{000}}\alpha _{1}^{3}-3a{{b}_{000}}{{\alpha }_{1}}\beta _{1}^{2}+c{{\alpha }_{1}}\beta _{1}^{2}+d\beta _{1}^{3}}{\beta _{1}^{2}} \right)t$,\\
${{\xi }_{3}}=\varepsilon {{\alpha }_{1}}x+\varepsilon {{\beta }_{1}}-\left( \frac{3a{{c}_{000}}\alpha _{1}^{3}-3a{{b}_{000}}{{\alpha }_{1}}\beta _{1}^{2}+c{{\alpha }_{1}}\beta _{1}^{2}+d\beta _{1}^{3}}{\beta _{1}^{2}} \right)t$,\\
(3) ${{\xi }_{1}}={{\alpha }_{1}}x+{{\beta }_{1}}y+\left( 3a{{b}_{000}}{{\alpha }_{1}}-a\alpha _{1}^{3}-c{{\alpha }_{1}}-d{{\beta }_{1}} \right)t$, ${{\xi }_{2}}=-\varepsilon I{{\beta }_{1}}y+\varepsilon Id{{\beta }_{1}}t$,
${{\xi }_{3}}=-{{\beta }_{1}}y+d{{\beta }_{1}}t$.

{\bf Case 10.} ${{d}_{2}}={{d}_{2}}$, ${{d}_{3}}=0$.\\
(1) ${{C}^{2}}=\frac{\left( {{\alpha }_{1}}+2{{\alpha }_{3}} \right)\left( {{\alpha }_{1}}-{{\alpha }_{3}} \right)}{{{\alpha }_{1}}\left( {{\alpha }_{1}}+{{\alpha }_{3}} \right)}$, ${{d}_{1}}={{d}_{2}}C$, ${{a}_{000}}=0$, ${{b}_{000}}={{b}_{000}}$, ${{c}_{000}}=\frac{4\beta _{3}^{2}}{3}$, ${{\alpha }_{1}}={{\alpha }_{1}}$,
${{\alpha }_{2}}=0$, ${{\alpha }_{3}}={{\alpha }_{3}}$, ${{\beta }_{1}}={{\beta }_{1}}$, ${{\beta }_{2}}=\varepsilon I{{\beta }_{1}}$, ${{\beta }_{3}}={{\beta }_{1}}$, ${{\gamma }_{1}}=3a{{b}_{000}}{{\alpha }_{1}}-a\alpha _{1}^{3}-\frac{3}{2}a\alpha _{1}^{2}{{\alpha }_{3}}-\frac{3}{2}a{{\alpha }_{1}}\alpha _{3}^{2}-c{{\alpha }_{1}}-d{{\beta }_{1}}$,
${{\gamma }_{2}}=\frac{\varepsilon I}{2}\left( 3a\alpha _{1}^{2}{{\alpha }_{3}}+3a{{\alpha }_{1}}\alpha _{3}^{2}-2d{{\beta }_{1}} \right)$,
${{\gamma }_{3}}=3a{{b}_{000}}{{\alpha }_{3}}-\frac{3}{2}a\alpha _{1}^{2}{{\alpha }_{3}}-\frac{3}{2}a{{\alpha }_{1}}\alpha _{3}^{2}-a\alpha _{3}^{3}-c{{\alpha }_{3}}-d{{\beta }_{1}}$.\\
(2) ${{C}^{2}}=\frac{\left( {{\alpha }_{1}}-2{{\alpha }_{3}} \right)\left( {{\alpha }_{1}}+{{\alpha }_{3}} \right)}{{{\alpha }_{1}}\left( {{\alpha }_{1}}-{{\alpha }_{3}} \right)}$, ${{d}_{1}}={{d}_{2}}C$, ${{a}_{000}}=0$, ${{b}_{000}}={{b}_{000}}$, ${{c}_{000}}=\frac{4\beta _{3}^{2}}{3}$, ${{\alpha }_{1}}={{\alpha }_{1}}$,
${{\alpha }_{2}}=0$, ${{\alpha }_{3}}={{\alpha }_{3}}$, ${{\beta }_{1}}={{\beta }_{1}}$, ${{\beta }_{2}}=-\varepsilon I{{\beta }_{1}}$, ${{\beta }_{3}}=-{{\beta }_{1}}$,
${{\gamma }_{1}}=3a{{b}_{000}}{{\alpha }_{1}}-a\alpha _{1}^{3}+\frac{3}{2}a\alpha _{1}^{2}{{\alpha }_{3}}-\frac{3}{2}a{{\alpha }_{1}}\alpha _{3}^{2}-c{{\alpha }_{1}}-d{{\beta }_{1}}$,
${{\gamma }_{2}}=\frac{\varepsilon I}{2}\left( 3a\alpha _{1}^{2}{{\alpha }_{3}}-3a{{\alpha }_{1}}\alpha _{3}^{2}+2d{{\beta }_{1}} \right)$,
${{\gamma }_{3}}=3a{{b}_{000}}{{\alpha }_{3}}-\frac{3}{2}a\alpha _{1}^{2}{{\alpha }_{3}}+\frac{3}{2}a{{\alpha }_{1}}\alpha _{3}^{2}-a\alpha _{3}^{2}-c{{\alpha }_{3}}+d{{\beta }_{1}}$.\\
Accordingly, we get ${{w}_{10}}={{e}^{-{{\xi }_{1}}}}+{{d}_{2}}C\cos {{\xi }_{2}}+{{d}_{2}}\cosh {{\xi }_{3}}$, where\\
(1) ${{\xi }_{1}}={{\alpha }_{1}}x+{{\beta }_{1}}y+\left( 3a{{b}_{000}}{{\alpha }_{1}}-a\alpha _{1}^{3}-\frac{3}{2}a\alpha _{1}^{2}{{\alpha }_{3}}-\frac{3}{2}a{{\alpha }_{1}}\alpha _{3}^{2}-c{{\alpha }_{1}}-d{{\beta }_{1}} \right)t$,\\
${{\xi }_{2}}=\varepsilon I{{\beta }_{1}}y+\frac{\varepsilon I}{2}\left( 3a\alpha _{1}^{2}{{\alpha }_{3}}+3a{{\alpha }_{1}}\alpha _{3}^{2}-2d{{\beta }_{1}} \right)t$,\\
${{\xi }_{3}}={{\alpha }_{3}}x+{{\beta }_{1}}y+\left( 3a{{b}_{000}}{{\alpha }_{3}}-\frac{3}{2}a\alpha _{1}^{2}{{\alpha }_{3}}-\frac{3}{2}a{{\alpha }_{1}}\alpha _{3}^{2}-a\alpha _{3}^{3}-c{{\alpha }_{3}}-d{{\beta }_{1}} \right)t$.\\
(2) ${{\xi }_{1}}={{\alpha }_{1}}x+{{\beta }_{1}}y+\left( 3a{{b}_{000}}{{\alpha }_{1}}-a\alpha _{1}^{3}+\frac{3}{2}a\alpha _{1}^{2}{{\alpha }_{3}}-\frac{3}{2}a{{\alpha }_{1}}\alpha _{3}^{2}-c{{\alpha }_{1}}-d{{\beta }_{1}} \right)t$,\\
${{\xi }_{2}}=-\varepsilon I{{\beta }_{1}}y+\frac{\varepsilon I}{2}\left( 3a\alpha _{1}^{2}{{\alpha }_{3}}-3a{{\alpha }_{1}}\alpha _{3}^{2}+2d{{\beta }_{1}} \right)t$,\\
${{\xi }_{3}}={{\alpha }_{3}}x-{{\beta }_{1}}y+\left( 3a{{b}_{000}}{{\alpha }_{3}}-\frac{3}{2}a\alpha _{1}^{2}{{\alpha }_{3}}+\frac{3}{2}a{{\alpha }_{1}}\alpha _{3}^{2}-a\alpha _{3}^{2}-c{{\alpha }_{3}}+d{{\beta }_{1}} \right)t$.

{\bf Case 11.} ${{d}_{1}}={{d}_{1}}$, ${{d}_{2}}={{d}_{2}}$, ${{d}_{3}}={{d}_{3}}$.\\
(1) ${{a}_{000}}=0$, ${{b}_{000}}={{b}_{000}}$, ${{c}_{000}}=-\frac{\beta _{2}^{2}}{3}$, ${{\alpha }_{1}}={{\alpha }_{1}}$, ${{\alpha }_{2}}=0$, ${{\alpha }_{3}}={{\alpha }_{3}}$, ${{\beta }_{1}}={{\beta }_{3}}=0$, ${{\beta }_{2}}={{\beta }_{2}}$, ${{\gamma }_{1}}=3a{{b}_{000}}{{\alpha }_{1}}-a\alpha _{1}^{3}-c{{\alpha }_{1}}$, ${{\gamma }_{2}}=-d{{\beta }_{2}}$, ${{\gamma }_{3}}=3a{{b}_{000}}{{\alpha }_{3}}-a\alpha _{3}^{3}-c{{\alpha }_{3}}$.\\
(2) ${{a}_{000}}=0$, ${{b}_{000}}={{b}_{000}}$, ${{c}_{000}}=\frac{\beta _{2}^{2}}{3}$, ${{\alpha }_{1}}=0$, ${{\alpha }_{2}}={{\alpha }_{2}}$, ${{\alpha }_{3}}={{\alpha }_{3}}$, ${{\beta }_{1}}={{\beta }_{1}}$, ${{\beta }_{2}}={{\beta }_{3}}=0$, ${{\gamma }_{1}}=-d{{\beta }_{1}}$, ${{\gamma }_{2}}={{\alpha }_{2}}\left( a\alpha _{2}^{2}+3a{{b}_{000}}-c \right)$, ${{\gamma }_{3}}=3a{{b}_{000}}{{\alpha }_{3}}-a\alpha _{3}^{3}-c{{\alpha }_{3}}$.\\
(3) ${{a}_{000}}=0$, ${{b}_{000}}={{b}_{000}}$, ${{c}_{000}}=-\frac{\beta _{2}^{2}}{3}$, ${{\alpha }_{1}}={{\alpha }_{2}}=0$, ${{\alpha }_{3}}={{\alpha }_{3}}$, ${{\beta }_{1}}=\varepsilon I{{\beta }_{2}}$, ${{\beta }_{2}}={{\beta }_{2}}$, ${{\beta }_{3}}=0$, ${{\gamma }_{1}}=-\varepsilon Id{{\beta }_{2}}$, ${{\gamma }_{2}}=-d{{\beta }_{2}}$, ${{\gamma }_{3}}=3a{{\alpha }_{3}}{{b}_{000}}-a\alpha _{3}^{3}-c{{\alpha }_{3}}$.\\
(4) ${{a}_{000}}=0$, ${{b}_{000}}={{b}_{000}}$, ${{c}_{000}}=\frac{\beta _{2}^{2}}{3}$, ${{\alpha }_{1}}=0$, ${{\alpha }_{2}}={{\alpha }_{2}}$, ${{\alpha }_{3}}=\varepsilon I{{\alpha }_{2}}$, ${{\beta }_{1}}=0$, ${{\beta }_{2}}=\varepsilon I{{\beta }_{3}}$, ${{\beta }_{3}}={{\beta }_{3}}$, ${{\gamma }_{1}}=0$, ${{\gamma }_{2}}=a\alpha _{2}^{3}+3a{{b}_{000}}{{\alpha }_{2}}-\varepsilon Id{{\beta }_{3}}-c{{\alpha }_{2}}$, ${{\gamma }_{3}}=\varepsilon Ia\alpha _{2}^{3}+3\varepsilon Ia{{b}_{000}}{{\alpha }_{2}}-\varepsilon Ic{{\alpha }_{2}}-d{{\beta }_{3}}$.\\
(5) ${{a}_{000}}=0$, ${{b}_{000}}={{b}_{000}}$, ${{c}_{000}}=\frac{\beta _{3}^{2}}{3}$, ${{\alpha }_{1}}={{\alpha }_{3}}=0$, ${{\alpha }_{2}}={{\alpha }_{2}}$, ${{\beta }_{1}}={{\beta }_{1}}$, ${{\beta }_{2}}=0$, ${{\beta }_{3}}=\varepsilon {{\beta }_{1}}$, ${{\gamma }_{1}}=\varepsilon d{{\beta }_{1}}$, ${{\gamma }_{2}}={{\alpha }_{2}}\left( a\alpha _{2}^{2}+3a{{b}_{000}}-c \right)$, ${{\gamma }_{3}}=-\varepsilon d{{\beta }_{1}}$.\\
(6) ${{a}_{000}}=0$, ${{b}_{000}}={{b}_{000}}$, ${{c}_{000}}=\frac{\beta _{3}^{2}}{3}$, ${{\alpha }_{1}}={{\alpha }_{1}}$, ${{\alpha }_{2}}={{\alpha }_{3}}=0$, ${{\beta }_{1}}=0$, ${{\beta }_{2}}=\varepsilon I{{\beta }_{3}}$, ${{\beta }_{3}}={{\beta }_{3}}$, ${{\gamma }_{1}}=-{{\alpha }_{1}}\left( a\alpha _{1}^{2}-3a{{b}_{000}}+c \right)$, ${{\gamma }_{2}}=-\varepsilon Id{{\beta }_{3}}$, ${{\gamma }_{3}}=-d{{\beta }_{3}}$.\\
(7) ${{a}_{000}}=0$, ${{b}_{000}}={{b}_{000}}$, ${{c}_{000}}=\frac{\beta _{3}^{2}}{3}$, ${{\alpha }_{1}}={{\alpha }_{1}}$, ${{\alpha }_{2}}={{\alpha }_{2}}$, ${{\alpha }_{3}}=0$, ${{\beta }_{1}}={{\beta }_{2}}=0$, ${{\beta }_{3}}={{\beta }_{3}}$, ${{\gamma }_{1}}=-{{\alpha }_{1}}\left( a\alpha _{1}^{2}-3a{{b}_{000}}+c \right)$, ${{\gamma }_{2}}={{\alpha }_{2}}\left( a\alpha _{2}^{2}+3a{{b}_{000}}-c \right)$, ${{\gamma }_{3}}=-d{{\beta }_{3}}$.\\
(8) ${{a}_{000}}=0$, ${{b}_{000}}={{b}_{000}}$, ${{c}_{000}}=\frac{4\beta _{3}^{2}}{3}$, ${{\alpha }_{1}}={{\alpha }_{1}}$, ${{\alpha }_{2}}=\varepsilon I{{\alpha }_{1}}$, ${{\alpha }_{3}}={{\alpha }_{1}}$, ${{\beta }_{1}}={{\beta }_{1}}$, ${{\beta }_{2}}=\varepsilon I{{\beta }_{1}}$, ${{\beta }_{3}}={{\beta }_{1}}$, ${{\gamma }_{1}}=3a{{b}_{000}}{{\alpha }_{1}}-4a\alpha _{1}^{3}-c{{\alpha }_{1}}-d{{\beta }_{1}}$,
${{\gamma }_{2}}=\varepsilon I\left( 3a{{b}_{000}}{{\alpha }_{1}}-4a\alpha _{1}^{3}-c{{\alpha }_{1}}-d{{\beta }_{1}} \right)$, ${{\gamma }_{3}}=3a{{\alpha }_{1}}{{b}_{000}}-4a\alpha _{1}^{3}-c{{\alpha }_{1}}-d{{\beta }_{1}}$.\\
(9) ${{a}_{000}}=0$, ${{b}_{000}}={{b}_{000}}$, ${{c}_{000}}=\frac{4\beta _{3}^{2}}{3}$, ${{\alpha }_{1}}={{\alpha }_{1}}$, ${{\alpha }_{2}}=-\varepsilon I{{\alpha }_{1}}$, ${{\alpha }_{3}}=-{{\alpha }_{1}}$, ${{\beta }_{1}}={{\beta }_{1}}$, ${{\beta }_{2}}=-\varepsilon I{{\beta }_{1}}$, ${{\beta }_{3}}=-{{\beta }_{1}}$, ${{\gamma }_{1}}=3a{{b}_{000}}{{\alpha }_{1}}-4a\alpha _{1}^{3}-c{{\alpha }_{1}}-d{{\beta }_{1}}$,
${{\gamma }_{2}}=\varepsilon I\left( 4a\alpha _{1}^{3}-3a{{b}_{000}}{{\alpha }_{1}}+c{{\alpha }_{1}}+d{{\beta }_{1}} \right)$, ${{\gamma }_{3}}=4a\alpha _{1}^{3}-3a{{b}_{000}}{{\alpha }_{1}}+c{{\alpha }_{1}}+d{{\beta }_{1}}$.\\
Accordingly, we get ${{w}_{11}}={{e}^{-{{\xi }_{1}}}}+{{d}_{1}}\cos {{\xi }_{2}}+{{d}_{2}}\cosh {{\xi }_{3}}+{{d}_{3}}{{e}^{{{\xi }_{1}}}}$, where\\
(1) ${{\xi }_{1}}={{\alpha }_{1}}x+\left( 3a{{b}_{000}}{{\alpha }_{1}}-a\alpha _{1}^{3}-c{{\alpha }_{1}} \right)t$, ${{\xi }_{2}}={{\beta }_{2}}y-d{{\beta }_{2}}t$,
${{\xi }_{3}}={{\alpha }_{3}}x+\left( 3a{{b}_{000}}{{\alpha }_{3}}-a\alpha _{3}^{3}-c{{\alpha }_{3}} \right)t$,\\
(2) ${{\xi }_{1}}={{\beta }_{1}}y-d{{\beta }_{1}}t$, ${{\xi }_{2}}={{\alpha }_{2}}x+{{\alpha }_{2}}\left( a\alpha _{2}^{2}+3a{{b}_{000}}-c \right)t$,
${{\xi }_{3}}={{\alpha }_{3}}x+\left( 3a{{b}_{000}}{{\alpha }_{3}}-a\alpha _{3}^{3}-c{{\alpha }_{3}} \right)t$,\\
(3) ${{\xi }_{1}}=\varepsilon I{{\beta }_{2}}y-\varepsilon Id{{\beta }_{2}}t$, ${{\xi }_{2}}={{\beta }_{2}}y-d{{\beta }_{2}}t$, ${{\xi }_{3}}={{\alpha }_{3}}x+\left( 3a{{\alpha }_{3}}{{b}_{000}}-a\alpha _{3}^{3}-c{{\alpha }_{3}} \right)t$,\\
(4) ${{\xi }_{1}}=0$, ${{\xi }_{2}}={{\alpha }_{2}}x+\varepsilon I{{\beta }_{3}}y+\left( a\alpha _{2}^{3}+3a{{b}_{000}}{{\alpha }_{2}}-\varepsilon Id{{\beta }_{3}}-c{{\alpha }_{2}} \right)t$,\\
${{\xi }_{3}}=\varepsilon I{{\alpha }_{2}}x+{{\beta }_{3}}y+\left( \varepsilon Ia\alpha _{2}^{3}+3\varepsilon Ia{{b}_{000}}{{\alpha }_{2}}-\varepsilon Ic{{\alpha }_{2}}-d{{\beta }_{3}} \right)t$,\\
(5) ${{\xi }_{1}}={{\beta }_{1}}y+\varepsilon d{{\beta }_{1}}t$, ${{\xi }_{2}}={{\alpha }_{2}}x+{{\alpha }_{2}}\left( a\alpha _{2}^{2}+3a{{b}_{000}}-c \right)t$, ${{\xi }_{3}}=\varepsilon {{\beta }_{1}}y-\varepsilon d{{\beta }_{1}}t$,\\
(6) ${{\xi }_{1}}={{\alpha }_{1}}x-{{\alpha }_{1}}\left( a\alpha _{1}^{2}-3a{{b}_{000}}+c \right)t$, ${{\xi }_{2}}=\varepsilon I{{\beta }_{3}}y-\varepsilon Id{{\beta }_{3}}t$, ${{\xi }_{3}}={{\beta }_{3}}y-d{{\beta }_{3}}t$,\\
(7) ${{\xi }_{1}}={{\alpha }_{1}}x-{{\alpha }_{1}}\left( a\alpha _{1}^{2}-3a{{b}_{000}}+c \right)t$, ${{\xi }_{2}}={{\alpha }_{2}}x+{{\alpha }_{2}}\left( a\alpha _{2}^{2}+3a{{b}_{000}}-c \right)t$,\\
${{\xi }_{3}}={{\beta }_{3}}y-d{{\beta }_{3}}t$,
(8) ${{\xi }_{1}}={{\alpha }_{1}}x+{{\beta }_{1}}y+\left( 3a{{b}_{000}}{{\alpha }_{1}}-4a\alpha _{1}^{3}-c{{\alpha }_{1}}-d{{\beta }_{1}} \right)t$,
${{\xi }_{2}}=\varepsilon I{{\alpha }_{1}}x+\varepsilon I{{\beta }_{1}}y+\varepsilon I\left( 3a{{b}_{000}}{{\alpha }_{1}}-4a\alpha _{1}^{3}-c{{\alpha }_{1}}-d{{\beta }_{1}} \right)t$,
${{\xi }_{3}}={{\alpha }_{1}}x+{{\beta }_{1}}y+\left( 3a{{\alpha }_{1}}{{b}_{000}}-4a\alpha _{1}^{3}-c{{\alpha }_{1}}-d{{\beta }_{1}} \right)t$,\\
(9) ${{\xi }_{1}}={{\alpha }_{1}}x+{{\beta }_{1}}y+\left( 3a{{b}_{000}}{{\alpha }_{1}}-4a\alpha _{1}^{3}-c{{\alpha }_{1}}-d{{\beta }_{1}} \right)t$,
${{\xi }_{2}}=-\varepsilon I{{\alpha }_{1}}x-\varepsilon I{{\beta }_{1}}y+\varepsilon I\left( 4a\alpha _{1}^{3}-3a{{b}_{000}}{{\alpha }_{1}}+c{{\alpha }_{1}}+d{{\beta }_{1}} \right)t$,
${{\xi }_{3}}=-{{\alpha }_{1}}x-{{\beta }_{1}}y+\left( 4a\alpha _{1}^{3}-3a{{b}_{000}}{{\alpha }_{1}}+c{{\alpha }_{1}}+d{{\beta }_{1}} \right)t$.

{\bf Case 12.} ${{d}_{1}}={{d}_{1}}$, ${{d}_{2}}={{d}_{2}}$.\\
(1) ${{d}_{3}}=\frac{d_{1}^{2}}{4}$,${{a}_{000}}=0$, ${{b}_{000}}={{b}_{000}}$, ${{c}_{000}}=-\frac{\beta _{2}^{2}}{3}$, ${{\alpha }_{1}}={{\alpha }_{1}}$, ${{\alpha }_{2}}=\varepsilon I{{\alpha }_{1}}$, ${{\alpha }_{3}}={{\alpha }_{3}}$, ${{\beta }_{1}}=\varepsilon I{{\beta }_{2}}$, ${{\beta }_{2}}={{\beta }_{2}}$, ${{\beta }_{3}}=0$, ${{\gamma }_{1}}=3a{{b}_{000}}{{\alpha }_{1}}-a\alpha _{1}^{3}-3a{{\alpha }_{1}}\alpha _{3}^{2}-c{{\alpha }_{1}}-\varepsilon Id{{\beta }_{2}}$,
${{\gamma }_{2}}=3\varepsilon Ia{{b}_{000}}{{\alpha }_{1}}-\varepsilon Ia\alpha _{1}^{3}-3\varepsilon Ia{{\alpha }_{1}}\alpha _{3}^{2}-\varepsilon Ic{{\alpha }_{1}}-d{{\beta }_{2}}$,
${{\gamma }_{3}}=3a{{b}_{000}}{{\alpha }_{3}}-3a\alpha _{1}^{2}{{\alpha }_{3}}-a\alpha _{3}^{3}-c{{\alpha }_{3}}$.\\
(2) ${{d}_{3}}=\frac{d_{2}^{2}}{4}$, ${{a}_{000}}=0$, ${{b}_{000}}={{b}_{000}}$, ${{c}_{000}}=\frac{\beta _{1}^{2}}{3}$, ${{\alpha }_{1}}={{\alpha }_{1}}$, ${{\alpha }_{2}}={{\alpha }_{2}}$, ${{\alpha }_{3}}=\varepsilon {{\alpha }_{1}}$, ${{\beta }_{1}}={{\beta }_{1}}$, ${{\beta }_{2}}=0$, ${{\beta }_{3}}=\varepsilon {{\beta }_{1}}$, ${{\gamma }_{1}}=3a{{\alpha }_{1}}\alpha _{2}^{2}-a\alpha _{1}^{3}+3a{{b}_{000}}{{\alpha }_{1}}-c{{\alpha }_{1}}-d{{\beta }_{1}}$,
${{\gamma }_{2}}=-{{\alpha }_{2}}\left( 3a\alpha _{1}^{2}-a\alpha _{2}^{2}-3a{{b}_{000}}+c \right)$, ${{\gamma }_{3}}=\varepsilon \left( a\alpha _{1}^{3}-3a{{\alpha }_{1}}\alpha _{2}^{2}-3a{{b}_{000}}{{\alpha }_{1}}+c{{\alpha }_{1}}-d{{\beta }_{1}} \right)$.\\
(3) ${{d}_{3}}=\frac{d_{2}^{2}}{4}$, ${{a}_{000}}=0$, ${{b}_{000}}={{b}_{000}}$, ${{c}_{000}}=\frac{\beta _{3}^{2}}{3}$, ${{\alpha }_{1}}={{\alpha }_{1}}$, ${{\alpha }_{2}}=0$, ${{\alpha }_{3}}=\varepsilon {{\alpha }_{1}}$, ${{\beta }_{1}}={{\beta }_{1}}$, ${{\beta }_{2}}=0$, ${{\beta }_{3}}=\varepsilon {{\beta }_{1}}$, ${{\gamma }_{1}}=a\alpha _{1}^{3}-3a{{b}_{000}}{{\alpha }_{1}}+c{{\alpha }_{1}}-d{{\beta }_{1}}$, ${{\gamma }_{2}}=0$,
${{\gamma }_{3}}=\varepsilon \left( 3a{{b}_{000}}{{\alpha }_{1}}-a\alpha _{1}^{3}-c{{\alpha }_{1}}-d{{\beta }_{1}} \right)$.\\
Accordingly, we get (1) ${{w}_{12}}={{e}^{-{{\xi }_{1}}}}+{{d}_{1}}\cos {{\xi }_{2}}+{{d}_{2}}\cosh {{\xi }_{3}}+\frac{d_{1}^{2}}{4}{{e}^{{{\xi }_{1}}}}$,\\
(2) ${{w}_{12}}={{e}^{-{{\xi }_{1}}}}+{{d}_{1}}\cos {{\xi }_{2}}+{{d}_{2}}\cosh {{\xi }_{3}}+\frac{d_{2}^{2}}{4}{{e}^{{{\xi }_{1}}}}$, (3) ${{w}_{12}}={{d}_{1}}+{{e}^{-{{\xi }_{1}}}}+{{d}_{2}}\cosh {{\xi }_{3}}+\frac{d_{2}^{2}}{4}{{e}^{{{\xi }_{1}}}}$, where\\
(1) ${{\xi }_{1}}={{\alpha }_{1}}x+\varepsilon I{{\beta }_{2}}y+\left( 3a{{b}_{000}}{{\alpha }_{1}}-a\alpha _{1}^{3}-3a{{\alpha }_{1}}\alpha _{3}^{2}-c{{\alpha }_{1}}-\varepsilon Id{{\beta }_{2}} \right)t$,\\
${{\xi }_{2}}=\varepsilon I{{\alpha }_{1}}x+{{\beta }_{2}}y+\left( 3\varepsilon Ia{{b}_{000}}{{\alpha }_{1}}-\varepsilon Ia\alpha _{1}^{3}-3\varepsilon Ia{{\alpha }_{1}}\alpha _{3}^{2}-\varepsilon Ic{{\alpha }_{1}}-d{{\beta }_{2}} \right)t$,\\
${{\xi }_{3}}={{\alpha }_{3}}x+\left( 3a{{b}_{000}}{{\alpha }_{3}}-3a\alpha _{1}^{2}{{\alpha }_{3}}-a\alpha _{3}^{3}-c{{\alpha }_{3}} \right)t$,\\
(2) ${{\xi }_{1}}={{\alpha }_{1}}x+{{\beta }_{1}}y+\left( 3a{{\alpha }_{1}}\alpha _{2}^{2}-a\alpha _{1}^{3}+3a{{b}_{000}}{{\alpha }_{1}}-c{{\alpha }_{1}}-d{{\beta }_{1}} \right)t$,\\
${{\xi }_{2}}={{\alpha }_{2}}x-{{\alpha }_{2}}\left( 3a\alpha _{1}^{2}-a\alpha _{2}^{2}-3a{{b}_{000}}+c \right)t$,\\
${{\xi }_{3}}=\varepsilon {{\alpha }_{1}}x+\varepsilon {{\beta }_{1}}y+\varepsilon \left( a\alpha _{1}^{3}-3a{{\alpha }_{1}}\alpha _{2}^{2}-3a{{b}_{000}}{{\alpha }_{1}}+c{{\alpha }_{1}}-d{{\beta }_{1}} \right)t$,\\
(3) ${{\xi }_{1}}={{\alpha }_{1}}x+{{\beta }_{1}}y+\left( a\alpha _{1}^{3}-3a{{b}_{000}}{{\alpha }_{1}}+c{{\alpha }_{1}}-d{{\beta }_{1}} \right)t$,\\
${{\xi }_{3}}=\varepsilon {{\alpha }_{1}}x+\varepsilon {{\beta }_{1}}y+\varepsilon \left( 3a{{b}_{000}}{{\alpha }_{1}}-a\alpha _{1}^{3}-c{{\alpha }_{1}}-d{{\beta }_{1}} \right)t$.

{\bf Case 13.}  ${{d}_{1}}={{d}_{1}}$, ${{d}_{2}}=-{{d}_{1}}$, ${{d}_{3}}={{d}_{3}}$.\\
${{a}_{000}}=0$, ${{b}_{000}}={{b}_{000}}$, ${{c}_{000}}=\frac{\beta _{3}^{2}}{3}$, ${{\alpha }_{1}}=0$, ${{\alpha }_{2}}={{\alpha }_{2}}$, ${{\alpha }_{3}}=\varepsilon I{{\alpha }_{2}}$, ${{\beta }_{1}}=0$, ${{\beta }_{2}}=\varepsilon I{{\beta }_{3}}$, ${{\beta }_{3}}={{\beta }_{3}}$, ${{\gamma }_{1}}=0$, ${{\gamma }_{2}}=a\alpha _{2}^{3}+3a{{b}_{000}}{{\alpha }_{2}}-\varepsilon Id{{\beta }_{3}}-c{{\alpha }_{2}}$,
${{\gamma }_{3}}=\varepsilon Ia\alpha _{2}^{3}+3\varepsilon Ia{{b}_{000}}{{\alpha }_{2}}-\varepsilon Ic{{\alpha }_{2}}-d{{\beta }_{3}}$.\\
Accordingly, we get ${{w}_{13}}={{d}_{1}}\cos {{\xi }_{2}}-{{d}_{1}}\cosh {{\xi }_{3}}+{{d}_{3}}$, where\\
${{\xi }_{2}}={{\alpha }_{2}}x+\varepsilon I{{\beta }_{3}}y+\left( a\alpha _{2}^{3}+3a{{b}_{000}}{{\alpha }_{2}}-\varepsilon Id{{\beta }_{3}}-c{{\alpha }_{2}} \right)t$,\\
${{\xi }_{3}}=\varepsilon I{{\alpha }_{2}}x+{{\beta }_{3}}y+\left( \varepsilon Ia\alpha _{2}^{3}+3\varepsilon Ia{{b}_{000}}{{\alpha }_{2}}-\varepsilon Ic{{\alpha }_{2}}-d{{\beta }_{3}} \right)t$.

{\bf Case 14.} ${{d}_{1}}={{d}_{1}}$, ${{d}_{2}}={{d}_{2}}$.\\
${{d}_{3}}=\frac{d_{1}^{2}\alpha _{1}^{2}{{\alpha }_{3}}-d_{2}^{2}\alpha _{1}^{2}{{\alpha }_{3}}+d_{1}^{2}{{\alpha }_{1}}\alpha _{2}^{2}-d_{2}^{2}{{\alpha }_{1}}\alpha _{2}^{2}+d_{1}^{2}{{\alpha }_{1}}\alpha _{3}^{2}-d_{2}^{2}{{\alpha }_{1}}\alpha _{3}^{2}+d_{1}^{2}\alpha _{2}^{2}{{\alpha }_{3}}}{\varepsilon I\left( 4{{\alpha }_{2}}\alpha _{3}^{2}-4\alpha _{1}^{2}{{\alpha }_{2}} \right)-8\alpha _{1}^{3}+4\alpha _{1}^{2}{{\alpha }_{3}}-4{{\alpha }_{1}}\alpha _{2}^{2}+4{{\alpha }_{1}}\alpha _{3}^{2}+4\alpha _{2}^{2}{{\alpha }_{3}}}+\\
\frac{d_{2}^{2}\alpha _{2}^{2}{{\alpha }_{3}}+2d_{2}^{2}\alpha _{3}^{3}+\varepsilon I\left( d_{1}^{2}\alpha _{1}^{2}{{\alpha }_{2}}-d_{2}^{2}\alpha _{1}^{2}{{\alpha }_{2}}+2d_{1}^{2}\alpha _{2}^{3}+d_{1}^{2}{{\alpha }_{2}}\alpha _{3}^{2}+d_{2}^{2}{{\alpha }_{2}}\alpha _{3}^{2} \right)}{\varepsilon I\left( 4{{\alpha }_{2}}\alpha _{3}^{2}-4\alpha _{1}^{2}{{\alpha }_{2}} \right)-8\alpha _{1}^{3}+4\alpha _{1}^{2}{{\alpha }_{3}}-4{{\alpha }_{1}}\alpha _{2}^{2}+4{{\alpha }_{1}}\alpha _{3}^{2}+4\alpha _{2}^{2}{{\alpha }_{3}}}, \\
$\\
${{a}_{000}}=0$, ${{b}_{000}}={{b}_{000}}$, ${{c}_{000}}=\frac{4\beta _{3}^{2}}{3}$, ${{\alpha }_{1}}={{\alpha }_{1}}$, ${{\alpha }_{2}}={{\alpha }_{2}}$, ${{\alpha }_{3}}={{\alpha }_{3}}$, ${{\beta }_{1}}={{\beta }_{1}}$, ${{\beta }_{2}}=\varepsilon I{{\beta }_{1}}$, ${{\beta }_{3}}={{\beta }_{1}}$,
${{\gamma }_{1}}=\frac{3\varepsilon I}{2}\left( a\alpha _{1}^{2}{{\alpha }_{2}}-a{{\alpha }_{2}}\alpha _{3}^{2} \right)+\frac{3}{2}\left( a{{\alpha }_{1}}\alpha _{2}^{2}-a\alpha _{1}^{2}{{\alpha }_{3}}-a{{\alpha }_{1}}\alpha _{3}^{2}-a\alpha _{2}^{2}{{\alpha }_{3}} \right)
  +3a{{b}_{000}}{{\alpha }_{1}}-c{{\alpha }_{1}}-d{{\beta }_{1}}-a\alpha _{1}^{3},
$
${{\gamma }_{2}}=a\alpha _{2}^{3}-\frac{3}{2}a\alpha _{1}^{2}{{\alpha }_{2}}-\frac{3}{2}a{{\alpha }_{2}}\alpha _{3}^{2}+3a{{b}_{000}}{{\alpha }_{2}}-c{{\alpha }_{2}}
  +\frac{3\varepsilon I}{2}\left( a\alpha _{1}^{2}{{\alpha }_{3}}+a{{\alpha }_{1}}\alpha _{2}^{2}+a{{\alpha }_{1}}\alpha _{3}^{2}+a\alpha _{2}^{2}{{\alpha }_{3}}-\frac{2}{3}d{{\beta }_{1}} \right),$
${{\gamma }_{3}}=\frac{3}{2}\left( a\alpha _{2}^{2}{{\alpha }_{3}}-a\alpha _{1}^{2}{{\alpha }_{3}}-a{{\alpha }_{1}}\alpha _{2}^{2}-a{{\alpha }_{1}}\alpha _{3}^{2} \right)-a\alpha _{3}^{3}
  +3a{{b}_{000}}{{\alpha }_{3}}-c{{\alpha }_{3}}-d{{\beta }_{1}}+\frac{3\varepsilon I}{2}\left( a{{\alpha }_{2}}\alpha _{3}^{2}-a\alpha _{1}^{2}{{\alpha }_{2}} \right). $
Accordingly, we get ${{w}_{14}}={{e}^{-{{\xi }_{1}}}}+{{d}_{1}}\cos {{\xi }_{2}}+{{d}_{2}}\cosh {{\xi }_{3}}+{{d}_{3}}{{e}^{{{\xi }_{1}}}}$, where\\
${{\xi }_{1}}={{\alpha }_{1}}x+{{\beta }_{1}}y+{{\gamma }_{1}}t$, ${{\xi }_{2}}={{\alpha }_{2}}x+I\varepsilon {{\beta }_{1}}y+{{\gamma }_{2}}t$, ${{\xi }_{3}}={{\alpha }_{3}}x+{{\beta }_{1}}y+{{\gamma }_{3}}t$.

Substituting ${{w}_{i}}\ \left( i=1,2,\cdots ,14 \right)$ into Eqs. (9), $u,\ v$ and $\omega $, the exact solutions of the GNNVEs, can be obtained.

{\bf Remark 3.} Choosing the appropriate parameters, the two-soliton solutions, the periodic solitary wave solutions, the doubly periodic solition solutions and the kink periodic two-soliton solutions of the GNNVEs can be obtained directly. For example, setting ${{d}_{3}}=1$ in case 4 yields ${{w}_{4}}=2\cosh {{\xi }_{1}}+{{d}_{2}}\cosh {{\xi }_{3}}$. Substituting ${{w}_{4}}$ into Eqs. (9), the two-soliton solutions of the GNNVEs can be obtained. Setting ${{d}_{3}}=1$ in case 7 yields ${{w}_{7}}=2\cosh {{\xi }_{1}}+{{d}_{1}}\cos {{\xi }_{2}}$. Substituting ${{w}_{7}}$ into Eqs. (9), the periodic solitary wave solutions of the GNNVEs can be obtained. Setting ${{\alpha }_{3}}=I{{\alpha }_{3}}$ in case 9-(1) yields ${{w}_{9}}={{e}^{-{{\xi }_{1}}}}+{{d}_{1}}\cos {{\xi }_{2}}+{{d}_{2}}\cos {{\xi }_{3}}$. Substituting ${{w}_{9}}$ into Eqs. (9), the doubly periodic solition solutions of the GNNVEs can be obtained. Setting ${{d}_{3}}=1$ in case 11 yields ${{w}_{11}}=\cosh {{\xi }_{1}}+{{d}_{1}}\cos {{\xi }_{2}}+{{d}_{2}}\cosh {{\xi }_{3}}$. Substituting ${{w}_{11}}$ into Eqs. (9), the kink periodic two-soliton solutions of the GNNVEs can be obtained.

{\bf Remark 4.} Similar to the three-wave method, by using the homoclinic test approach [34], we can get some exact solutions of the GNNVEs. It being a similar process, we omit it.

{\bf Remark 5.} Setting $c=d=0$, the results can apply to the NNVEs.
\section{Conclusions}
The GNNVEs are converted into the combined equations of differently two bilinear forms by means of the homogeneous balance of undetermined coefficients method. Then, the two class of exact $N$-soliton solutions and three wave solutions, which are influence of initial solutions $u={{a}_{000}},v={{b}_{000}},\omega ={{c}_{000}}$, are obtained respectively by using the Hirota's direct method and the three wave method.

This result shows that the method proposed in the paper is effective for NLPDEs which can't be transformed to complete form of bilinear operator. The performance of this method is found to be simple and efficient. The availability of computer systems like Maple facilitates the tedious algebraic calculations. The proposed method is also a standard and computable method, which can be generalized to deal with some NLPDEs.

\vskip 0.4 true cm {\bf Acknowledgments }
The research is supported by the National Natural Science Foundation of China (11372252, 11372253 and 11432010), the Fundamental Research Funds for the Central Universities (3102014JCQ01035).\\ \\
\vskip 0.4 true cm



\bigskip
\bigskip

\end{document}